\documentclass[fleqn,usenatbib]{mnras}

\usepackage{amssymb}

\usepackage{newtxtext,newtxmath}
\usepackage[T1]{fontenc}
\DeclareRobustCommand{\VAN}[3]{#2}
\let\VANthebibliography\thebibliography
\def\thebibliography{\DeclareRobustCommand{\VAN}[3]{##3}\VANthebibliography}

\usepackage{graphicx}   % Including figure files
\usepackage{amsmath}    % Advanced maths commands
\usepackage{bm}
\usepackage{hyperref}

\defcitealias{L21}{Paper I}
\defcitealias{KT18}{KT18}

\newcommand{\aref}[1]{\hyperref[#1]{Appendix \ref{#1}}}

\usepackage{scalerel,tikz}
\usetikzlibrary{svg.path}
\definecolor{orcidlogocol}{HTML}{A6CE39}
\tikzset{orcidlogo/.pic={
 \fill[orcidlogocol] svg{M256,128c0,70.7-57.3,128-128,128C57.3,256,0,198.7,0,128C0,57.3,57.3,0,128,0C198.7,0,256,57.3,256,128z};
 \fill[white] svg{M86.3,186.2H70.9V79.1h15.4v48.4V186.2z}
 svg{M108.9,79.1h41.6c39.6,0,57,28.3,57,53.6c0,27.5-21.5,53.6-56.8,53.6h-41.8V79.1z M124.3,172.4h24.5c34.9,0,42.9-26.5,42.9-39.7c0-21.5-13.7-39.7-43.7-39.7h-23.7V172.4z}
 svg{M88.7,56.8c0,5.5-4.5,10.1-10.1,10.1c-5.6,0-10.1-4.6-10.1-10.1c0-5.6,4.5-10.1,10.1-10.1C84.2,46.7,88.7,51.3,88.7,56.8z};
}}
\newcommand\orcidicon[1]{\href{https://orcid.org/#1}{\mbox{\scalerel*{
\begin{tikzpicture}[yscale=-1,transform shape]
\pic{orcidlogo};
\end{tikzpicture}
}{|}}}}

\title[Spatial metallicity distributions in AMUSING++]{Spatial metallicity distribution statistics at $\lesssim 100$ pc scales in the AMUSING++ nearby galaxy sample}

\author[Z. Li et al.]{Zefeng Li$^{\orcidicon{0000-0001-7373-3115}}$, $^{1, 2}$\thanks{E-mail: zefeng.li@anu.edu.au}
Emily Wisnioski$^{\orcidicon{0000-0003-1657-7878}}$, $^{1, 2}$
J. Trevor Mendel$^{\orcidicon{0000-0002-6327-9147}}$, $^{1, 2}$
Mark R. Krumholz$^{\orcidicon{0000-0003-3893-854X}}$, $^{1, 2}$
\newauthor Lisa J. Kewley$^{\orcidicon{0000-0001-8152-3943}}$, $^{1, 2}$
Carlos L\'opez-Cob\'a$^{\orcidicon{0000-0003-1045-0702}}$, $^3$
Sebastian F. S\'anchez$^{\orcidicon{0000-0001-6444-9307}}$, $^4$ Joseph P.
\newauthor Anderson$^{\orcidicon{0000-0003-0227-3451}\, 5}$
 and Llu\'is Galbany$^{\orcidicon{0000-0002-1296-6887}\, 6, 7}$ \\
$^1$Research School of Astronomy \& Astrophysics, Australian National University, Weston Creek, ACT 2611, Australia\\
$^2$ARC Centre of Excellence for All Sky Astrophysics in 3 Dimensions (ASTRO 3D), Canberra, ACT 2611, Australia\\
$^3$Institute of Astronomy \& Astrophysics, Academia Sinica, Taipei, 10617, Taiwan\\
$^4$Instituto de Astronom\'ia, Universidad Nacional Auton\'oma de M\'exico, A.P. 70-264, 04510 M\'exico, D.F., Mexico\\
$^5$European Southern Observatory, Alonso de C\'ordova 3107, Vitacura, Casilla 190001, Santiago, Chile\\
$^6$Institute of Space Sciences (ICE, CSIC), Campus UAB, Carrer de Can Magrans, s/n, E-08193 Barcelona, Spain\\
$^7$Institut d’Estudis Espacials de Catalunya (IEEC), E-08034 Barcelona, Spain
}

\date{Accepted XXX. Received YYY; in original form ZZZ}

\pubyear{2015}

\begin{document}
\label{firstpage}
\pagerange{\pageref{firstpage}--\pageref{lastpage}}
\maketitle

% Abstract of the paper
\begin{abstract}
We analyse the spatial statistics of the 2D gas-phase oxygen abundance distributions in a sample of 219 local galaxies. We introduce a new adaptive binning technique to enhance the signal to noise ratio of weak lines, which we use to produce well-filled metallicity maps for these galaxies. We show that the two-point correlation functions computed from the metallicity distributions after removing radial gradients are in most cases well described by a simple injection–diffusion model. Fitting the data to this model yields the correlation length $l_{\rm corr}$, which describes the characteristic interstellar medium mixing length scale. We find typical correlation lengths $l_{\rm corr} \sim 1$ kpc, with a strong correlation between $l_{\rm corr}$ and stellar mass, star formation rate, and effective radius, a weak correlation with Hubble type, and significantly elevated values of $l_{\rm corr}$ in interacting or merging galaxies. We show that the trend with star formation rate can be reproduced by a simple transport+feedback model of interstellar medium turbulence at high star formation rate, and plausibly also at low star formation rate if dwarf galaxy winds have large mass-loading factors. We also report the first measurements of the injection width that describes the initial radii over which supernova remnants deposit metals. Inside this radius the metallicity correlation function is not purely the product of a competition between injection and diffusion. We show that this size scale is generally smaller than 60 pc.
\end{abstract}

% Select between one and six entries from the list of approved keywords.
% Don't make up new ones.

%%%%%%%%%%%%%%%%%%%%%%%%%%%%%%%%%%%%%%%%%%%%%%%%%%

%%%%%%%%%%%%%%%%% BODY OF PAPER %%%%%%%%%%%%%%%%%%

\section{Introduction}

Metals (chemical elements heavier than helium) are forged inside stars and redistributed when stars reach the ends of their lives. Once ejected into the surrounding interstellar medium (ISM), some of the metals will be incorporated into the next generation of stars. This cycle makes metals in both gaseous and stellar phases natural tracers of galactic and chemical evolution \citep[for reviews, see][]{Tinsley80, Maiolino19, Sanchez21}.

Measurements of the gas-phase oxygen abundances (hereafter metallicities) of H~\textsc{ii} regions can be made using both direct electron temperature methods and strong emission-line diagnostics \citep[for a review, see][]{Kewley19}. Traditional spectrographs used these tools to explore the global metallicities of nearby galaxies \citep[e.g.][]{Tremonti04, Gallazzi05}, and in some cases metallicity gradients using long-slit spectra \citep[e.g.][]{Vila-Costas92, Henry99}. The development of integral field spectroscopy (IFS) surveys \citep[e.g.][]{Marmol-Queralto11, Croom12, Sanchez12, Bundy15, Erroz-Ferrer19, Lopez-Coba20, Emsellem21} has recently enabled studies of the full 2D spatially-resolved metallicities of nearby galaxies \citep[e.g.][]{Rosales-Ortega11, Sanchez14, Sanchez-Menguiano16_conf, Sanchez-Menguiano20, Grasha22, Metha22}. Large IFS surveys have allowed statistical studies of metallicity gradients which confirm earlier results that nearby galaxies generally have a negative azimuthally-averaged metallicity gradient - metallicities in central regions are higher than outer ones \citep[e.g.][]{Belfiore17, Poetrodjojo18, Sanchez-Menguiano18, Kreckel19}. Full 2D metallicity maps have also made it possible to study azimuthal variations \citep[e.g.][]{Sanchez-Menguiano16_azi, Ho18, Sanchez-Menguiano20_arm}.

Driven by the goal of extracting more information from metallicity maps, \citet[][hereafter \citetalias{KT18}]{KT18} proposed a first-principles model based on stochastically forced diffusion that predicts the two-point correlation functions of metallicity fields caused by the competition between chemical mixing and metal production. Following \citetalias{KT18} a number of recent studies have investigated the statistical correlations of local galaxies. \citet[][hereafter \citetalias{L21}]{L21} apply the \citetalias{KT18} model to a hundred galaxies from the Calar Alto Legacy Integral Field spectroscopy Area (CALIFA) survey, and report that the correlation scale of the metallicity, which is correlated with the rate of metal mixing in the ISM, positively correlates with several galactic properties including stellar mass ($M_*$), star formation rate (SFR), and effective radius ($R_e$). \cite{Kreckel20} and \cite{Williams22} analyse two-point correlations of 8 PHANGS-MUSE nearby galaxies. \cite{Metha21} use the semivariogram, a mathematical tool similar to the two-point correlation function, to analyse the size and spatial scale of metallicity fluctuations of 7 PHANGS-MUSE galaxies.

These studies push the frontier of metallicity map analysis to an unprecedented level - metallicity fluctuations on top of well-studied metallicity gradients can provide crucial constraints on models of the ISM and galaxy formation. Compared to our understanding of the nucleosynthesic processes that produce metals, we have poor knowledge of how metals are transported once they are created. It still remains an open question what typical ISM metal transport length / time scales are, and how they are influenced by other galactic-scale activities, e.g. gas inflow / outflow. Observational constraints therefore serve as an important to guide theoretical models \citep[e.g.][]{Sharda21} and numerical simulations \citep[e.g.][]{deAvillez02, Kobayashi11, Yang12, Minchev13, Petit16, Ceverino16, Colbrook17, Escala18} of star formation history and galaxy chemical evolution.

However, the information that can be gleaned from the observational analyses published to date is limited, because existing samples suffer from either low spatial resolution or small sample size. PHANGS-MUSE achieves $\sim 50$ pc resolution, but in currently $<20$ galaxies, and covering a narrow range of galaxy mass and morphological type. This makes it difficult to use these data to examine relationships between metal transport and other galaxy properties. Conversely, \citetalias{L21} samples an order of magnitude more galaxies, but with a resolution of several hundred pc. This is sufficient to measure gross characteristics of the metals such as the correlation length, and its relationship with other galaxy properties, but the spatial resolution is not enough to look at finer details of the metallicity distribution.

The goal of this paper is to extend the analysis techniques developed in \citetalias{L21}, which build on the basic statistical tool proposed in \citetalias{KT18}, to both higher spatial resolution and larger sample sizes. The former makes it possible for the first time to constrain details in the shape of the metallicity correlation, which potentially constrain the details of metal injection as well as transport through the ISM. The latter, on the other hand, can extend our scope, especially to dwarf galaxies with low $M_*$ and SFR. For this purpose, we analyse galaxies drawn from the AMUSING++ compilation \citep{Lopez-Coba20}, which provides much higher spatial resolution than CALIFA, but a much larger and more diverse galaxy sample than PHANGS-MUSE.

The outline of this paper is as follows. In \autoref{sec:data}, we provide an overview of the galaxy catalogue, our selection criteria, and the method we use to derive metallicities from the observations. In \autoref{sec:method}, we discuss our method for analysing the spatial metallicity distributions and finalise the sample selection; part of this method involves a novel binning algorithm for the purpose of signal-to-noise ratio enhancement. In \autoref{sec:results}, we describe the two main results from the pipeline, galactic correlation lengths and injection widths, which are the two fundamental quantities in our model for the two-point correlation function. In \autoref{sec:discussion} we discuss the potential insight provided by the extracted correlation lengths and compare them with previous work. Finally, we draw conclusions in \autoref{sec:conclusions}.

\section{Data}
\label{sec:data}

The instrument Multi Unit Spectroscopic Explorer \citep[MUSE;][]{Bacon10} is an integral field unit (IFU) at the Very Large Telescope (VLT). It has a field of view (FoV) of $1'\times1'$, and a spatial sampling of $0\farcs2\times0\farcs2$. MUSE has a wavelength coverage from 4650 to 9300 \AA, and achieves a spectral resolution of 1750 (at 4650 \AA) and 3750 (at 9300 \AA). The combined spectral and spatial resolution provides unique opportunities to explore the elemental abundance distribution in galaxies \citep[e.g.][]{Sanchez-Menguiano18}.

AMUSING++ \citep{Lopez-Coba20} is currently the largest compilation of nearby galaxies observed by MUSE. It consists of 532 galaxies from several different surveys \citep{Lopez-Coba20}; a majority of these come from the All-weather MUse Supernova Integral-field of Nearby Galaxies \citep[AMUSING;][]{Galbany16}. The emission lines and the stellar population content of the data cubes were derived using \textsc{pipe3d} \citep{Sanchez16_pipe3d}, which is a fitting routine adapted to analyse IFS data using the package \textsc{fit3d} \footnote{\href{http://ifs.astroscu.unam.mx/pyPipe3D/}{http://ifs.astroscu.unam.mx/pyPipe3D/}} \citep{Sanchez16_fit3d}.

We describe the procedure to obtain the full widths at half maximum (FWHM) of the point spread functions (PSFs) of the AMUSING++ sample in \aref{app:psf}. The median FWHM for the AMUSING++ galaxies is $0\farcs91$. The median distance of galaxies in the AMUSING++ sample is similar to that of CALIFA ($\sim 70$ Mpc), but since AMUSING++ has a much smaller angular PSF than CALIFA ($\sim 1''$ versus $\sim 2\farcs5$), it achieves much finer physical resolution ($\sim 300$ pc versus $\sim 800$ pc). Compared with PHANGS-MUSE, AMUSING++ uses the same instrument and has similar seeing conditions. The closer distances (median $\sim 11$ Mpc) of the PHANGS-MUSE sample make its physical resolution even finer ($\sim 50$ pc), but at the price of a much smaller sample size covering a more limited range of galaxy types.

\begin{figure}
\includegraphics[width=1.0\linewidth]{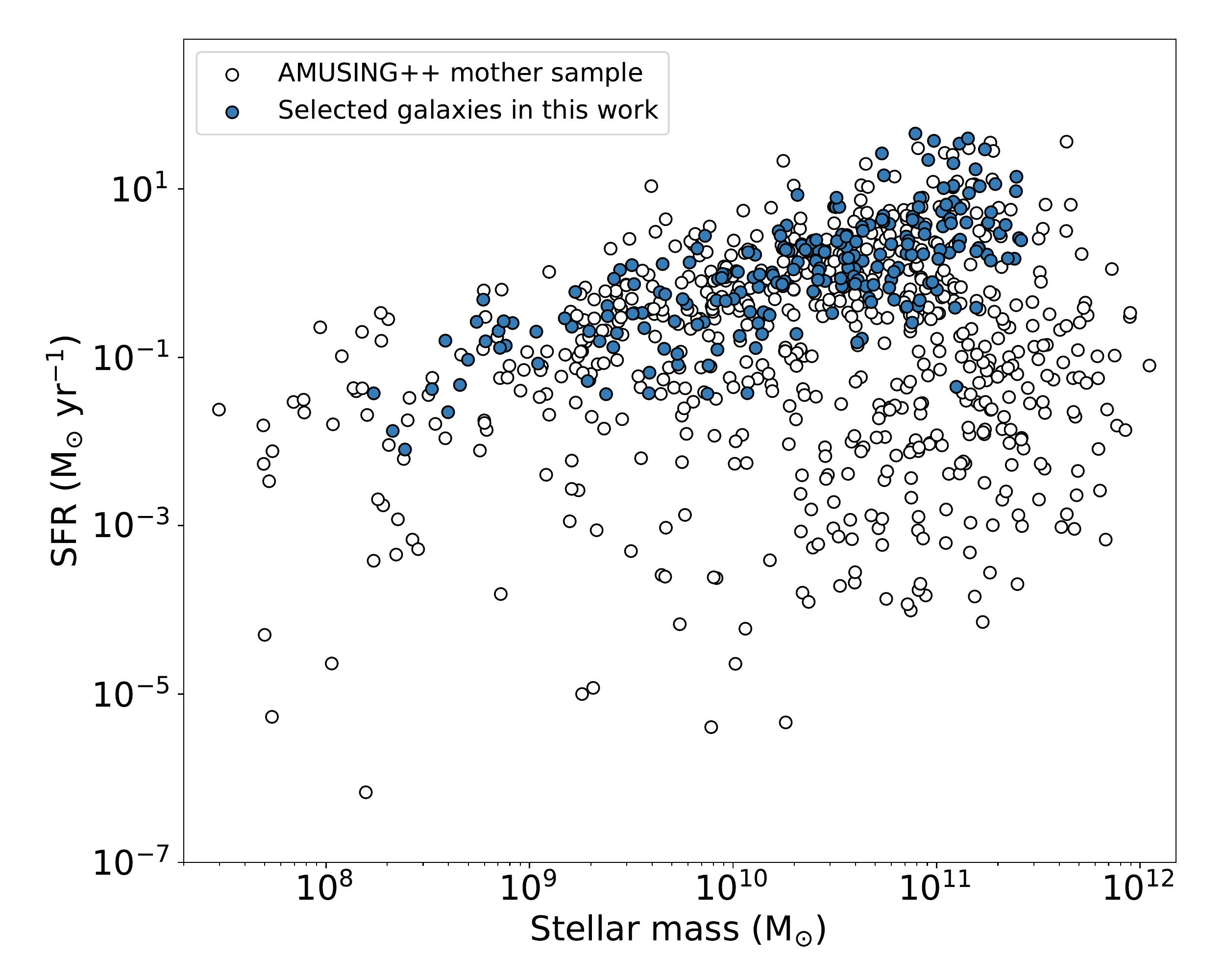}
\caption{Left panel: Dust-corrected H$\alpha$ star formation rate versus stellar mass (integrated value estimated by Pipe3D). The open circles show the full AMUSING++ spectroscopic sample of 532 galaxies, while blue circles show the sample we select for analysis in this work (219 galaxies, see \autoref{subsec:selection}). Right panel: number distribution of the selected galaxies. The median value is $0\farcs91$.}
\label{fig:sample}
\end{figure}

We derive gas-phase metallicities for AMUSING++ galaxies using the available strong emission lines. While in \citetalias{L21} we use the [N \textsc{ii}]/[O \textsc{ii}] diagnostic \citep{Kewley19} due to its insensitivity to ionisation parameter, the wavelength coverage of MUSE does not include [O \textsc{ii}] for nearby galaxies. We therefore use the metallicity diagnostic proposed by \citet[][abbreviated as D16]{Dopita16}, 
\begin{eqnarray}
12 + \log(\rm O/\rm H) & = & 8.77 + y + 0.45(y + 0.3)^5,
\label{eqn:D16} \\
y & = & \log([\rm N \textsc{ii}]\lambda6584/[\rm S \textsc{ii}]\lambda\lambda6717,31) \nonumber \\
& & \qquad {} + 0.264\log([\rm N \textsc{ii}]\lambda6584/\rm H\textsc{$\alpha$}).
\end{eqnarray}
D16 shows the least dependence on the ionisation parameter among the current optical metallicity diagnostics \citep[$U$;][]{Kewley19}. In principle one can estimate $\log(U)$ using the existing strong lines, but we find that attempting to do so yields unreliable results, we therefore adopt the original D16 diagnostic without making a separate estimate of $\log(U)$; we discuss this issue in detail in \aref{app:ion_par}. Comparison of the results from different metallicity diagnostics is also illustrated in \aref{app:diag}, where we show that using a different diagnostic does not change our qualitative results. Before inferring the metallicity from any diagnostic, we correct the AMUSING++ line fluxes for the extinction within the source galaxies using the $E(B-V)$ map provided as part of the AMUSING++ compilation, assuming  the extinction curve proposed by \cite{Calzetti00}. We do not correct for Galactic extinction, as that correction (computed using the \citealt{Cardelli89} model) is already included in the line fluxes reported in the AMUSING++ data products.

To measure reliable correlation functions for the 2D metal field, we require galaxies to have an axis ratio $b/a > 0.4$. The requirement limits the sample to galaxies that can be deprojected accurately and downsizes the whole sample from 532 to 447 galaxies. Next, three criteria are used to mask pixels where the ionisation is not primarily due to star formation, or shows significant contamination by shocks or AGN; we cannot reliably estimate the metallicity in these pixels using the D16 diagnostic. First, we mask non-star-forming pixels using the criterion proposed by \citet{Kauffmann03}. This represents a more conservative masking than if we were to use the AGN line proposed by \citet{Kewley01}, but we make this more conservative choice because the D16 diagnostic is relatively sensitive to shock and AGN contamination, with even a 20\% contribution yielding appreciable errors \citep{KE08}. Second, we mask the pixels where the equivalent width (EW) of H\textsc{$\alpha$} is $<6$ \AA, because ionisation in low EW regions is usually dominated by old stellar populations \citep[e.g.][]{Sanchez14, Sanchez15, Espinosa-Ponce20}; the D16 diagnostic is also unreliable for these regions. Third, we mask the pixels for which the $12 + \log(\mathrm{O}/\mathrm{H})$ value returned by the D16 diagnostic is higher than 9.23 or lower than 7.63 \citep{Kewley19}, meaning that it is outside the range where D16 provides valid results. The pixel removal reduces the coverage of the metallicity map, and this renders some of the maps too sparse to provide useful constraints on metallicity correlations. We provide a full description of our treatment of missing pixels in \autoref{subsec:selection} after discussing the rest of our analysis pipeline, and in that Section we determine criteria for galaxies having sufficient coverage to allow accurate analysis. After applying all these criteria, we are left with a sample of 219 galaxies out of the 447 with which we started.

In the left panel of \autoref{fig:sample} we show the distribution of star formation rate and stellar mass for both the full AMUSING++ sample and the sub-sample of 219 galaxies that pass all our selection criteria. We see that the effect of our selection is primarily to remove galaxies with low star formation rates for their stellar masses, which sit below the star-forming main sequence. Not surprisingly, such galaxies tend to have relatively sparse coverage of H~\textsc{ii} regions that allow reliable metallicity inference. However, our sample still covers a broad range of masses for galaxies on the main sequence.

\section{Analysis method}
\label{sec:method}

As discussed previously, this work is focused on analysis of the two-dimensional distribution of metals via the two-point correlation function. The backbone of the analysis method generally follows \citetalias{L21}, though we make some changes that we describe in this section. We first introduce an adaptive binning scheme in \autoref{subsec:abr}. In \autoref{subsec:backbone} we summarise our pipeline, including metallicity maps, two-point correlation functions, and the parametric model fit; the steps here are similar to those in \citetalias{L21}, and we refer readers there for full details. In \autoref{subsec:selection} we discuss our selection criteria, which downsize the full AMUSING++ mother sample to a subset for which we can conduct the analysis reliably. In addition, we have performed an extensive series of tests on our analysis pipeline in order to verify that the results we derive are robust against various choices of parameters, choices of metallicity diagnostics (Section 4.1 of \citetalias{L21}), and analysis method, and against variations in the observing conditions. We describe these validation tests in \aref{app:test}.

\subsection{Adaptive binning reconstruction}
\label{subsec:abr}

\begin{figure*}
\includegraphics[width=1.0\linewidth]{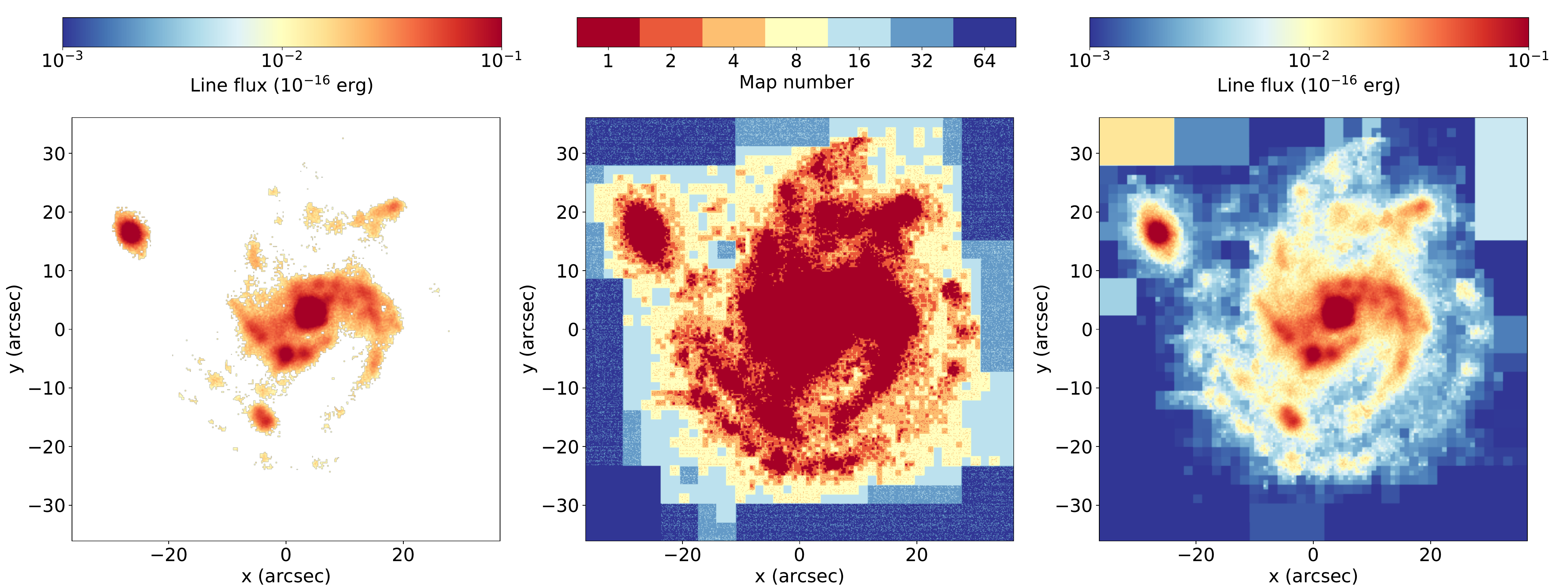}
\caption{An illustration of adaptive binning reconstruction, using NGC 7674 (the host galaxy of SN2011hb) with a target S/N ratio of 3 as an example. The left panel shows the original [S \textsc{ii}]$\lambda6731$ map; blank areas show masked pixels with S/N $<3$. The middle panel shows at which map (map$_N$) each pixel reaches the target S/N. Note that red area in the middle panel corresponds exactly to the non-masked pixels in the left panel, and these pixels will not be altered by adaptive binning because they are already above the target S/N. The right panel shows the final, adaptively binned [S \textsc{ii}]$\lambda6731$ map.}
\label{fig:recon}
\end{figure*}

Of the four emission lines on which the D16 metallicity diagnostic relies (H\textsc{$\alpha$}, [N \textsc{ii}]$\lambda6584$, [S \textsc{ii}]$\lambda6717,31$), [S \textsc{ii}]$\lambda6731$ has the lowest signal-to-noise ratio (S/N) and H\textsc{$\alpha$} the highest. If we simply require a minimum S/N of 3 in each line in order to derive a metallicity for a given pixel, the number of pixels that meet this threshold in the [S \textsc{ii}]$\lambda6731$ line is typically 30\% of the number that do so in the H\textsc{$\alpha$}. This makes [S \textsc{ii}]$\lambda6731$ a bottleneck, since the number of pixels for which we can estimate metallicities from \autoref{eqn:D16} is determined by the weakest line. Since we are interested in spatial statistics, having large gaps in our map where the signal is too weak to measure the metallicity presents obvious problems.

In order to recover as many spatial pixels as possible, we use an adaptive binning scheme.\footnote{The source code for our method is available at \\
\href{https://doi.org/10.5281/zenodo.6517216}{https://doi.org/10.5281/zenodo.6517216}.} For independent, identically distributed data we expect the S/N of binned data to increase in proportion to the square root of the number of binned pixels. This increase is paid for by a corresponding decrease in spatial resolution, and thus we only want to bin pixels where we are required to do so for S/N reasons. We therefore carry out the following steps:
\begin{enumerate}
    \item First we create a series of maps which cover the same area as our data, which we denote map$_1$, map$_2$, map$_4$, map$_8$, and so forth, where map$_N$ is a map where we have binned each group of $N^2$ adjacent pixels in our data together. Therefore map$_1$ is our original data, while in map$_2$ we have averaged together every $2\times 2$ set of pixels into a single pixel; thus map$_2$ has half the resolution and $4\times$ fewer pixels than map$_1$, but each pixel has $2\times$ the S/N of the original data. Similarly, in map$_4$ we bin together every set of $4\times 4$ pixels from the original map to create a map with $16\times$ fewer pixels but $4\times$ the S/N, and so forth.
    \item We then generate an output map with the same size as map$_1$. For each pixel in the output map, we first locate the corresponding pixel in map$_1$, and ask if its S/N ratio is above some specified threshold. If so, we set the value of the pixel in the output map to the value of the pixel in map$_1$. If the S/N does not reach our threshold, we then examine the pixel in map$_2$ that covers the same area, and use its value instead if the S/N ratio is high enough. If not, we proceed to map$_4$, and so forth. We fill in every pixel in the output map in this manner.
\end{enumerate}

We illustrate this process in \autoref{fig:recon}. The left panel shows the original data, with pixels below a S/N ratio of 3 masked. The middle panel shows the value $N$ of the map$_N$ for which the S/N reaches the target S/N of 3. As is clear from the figure, the algorithm uses high-resolution data in high S/N regions, and degrades smoothly to more and more binned data in regions of weak signal. The right panel then shows the final output, adaptively binned map. Again, we see that high-resolution information has been preserved where possible, but now none of the map area is masked.

We apply the adaptive binning scheme to the [S \textsc{ii}]$\lambda6731$ line map first, using a S/N ratio target of 3. We choose this value because in tests of our pipeline we have found that the choice of the target S/N ratio does not significantly influence the results (see \aref{app:sn}), and we therefore adopt a S/N ratio of 3. For consistency we then reconstruct the fluxes in all other lines using the same binning, i.e., using the map number shown in the middle panel of \autoref{fig:recon}.  We do this in order to ensure that we only ever compute metallicities using ratios of lines that are measured as the same spatial resolution. In the final, adaptively binned maps, we then mask any pixels where \textit{none} of the lines in the original, unbinned map were detected at S/N $>3$; in practice, since H$\alpha$ is almost always the brightest line, this means that our final, adaptively binned maps are limited to the footprint where H$\alpha$ was detected in the original, unbinned map. This ensures that our adaptively binned maps do not go past the real edge of ionised gas emission in the galaxy.

We also investigated other possible binning schemes, e.g. Voronoi binning \citep{Cappellari03}. However, we find that the adaptive binning scheme described above offers superior performance for our data because it is better able to handle the non-axisymmetric features of star-forming galaxies. Although the Voronoi binning algorithm parameter ``roundness'' accounts for the shape of bins, the algorithm is optimised for elliptical galaxies, and applying it to our data both leaves large areas that do not reach the target S/N, and leads to the production of binned Voronoi pixels whose shapes leave obvious artefacts when we compute metallicity maps and derive spatial statistics from them.

\subsection{The backbone: metallicity maps, two-point correlation functions, and the parametric model fit}
\label{subsec:backbone}

Once we have adaptively binned the data, we run the rest of our pipeline, which is substantially the same as in \citetalias{L21}. The pipeline starts by rotating the images to align the major axis of the original image to the $x$-axis, and then deprojecting the galaxy from an ellipse to a circle. A rotation matrix does both at the same time, converting the original coordinate of each pixel ($x, y$) to a new one ($x', y'$) as
\begin{equation}
\left[\begin{array}{c}
x'\\
y'
\end{array}\right]
=
\left[\begin{array}{cc}
\cos\theta & \sin\theta \\
-\sin\theta/\cos i & \cos\theta/\cos i
\end{array}\right]
\left[\begin{array}{c}
x\\
y
\end{array}\right],
\end{equation}
where $\theta$ represents the position angle (PA) and $i$ is the inclination angle. We determine $i$ using the classical Hubble formula \citep[][]{Hubble26}
\begin{equation}
\cos^2 i = \frac{(b/a)^2-q_0^2}{1-q_0^2},
\label{eqn:cosi}
\end{equation}
where $b/a$ is the axis ratio in the original image and $q_0 = 0.13$ \citep[][$i=90^{\circ}$ for $b/a<q_0$]{Giovanelli94}. We adopt 
%a $q_0=0.13$ 
this value of $q_0$
although the sample consists of a wide range of Hubble types (see \autoref{subsec:lcorr} and \autoref{fig:morph}). Adopting a larger $q_0$ does not change our results (see \aref{app:q0}).

After deprojecting each of the line maps, we generate a deprojected metallicity map (each value denoted as $Z_i$) using the D16 diagnostic. The next step is to remove the radial metallicity gradient to produce a fluctuation map. We choose to subtract the mean value, $\overline{Z_r}$, computed in annular bins of a bin width which is the same as the physical sampling scale ($0\farcs2$). The physical bin width varies from galaxy to galaxy; the median value over the sample is 64 pc. The metallicity fluctuation $Z'_i$ for each pixel is $Z'_i = Z_i - \overline{Z_r}$. This fluctuation map by construction has zero mean.

We now compute the two-point correlation function of this metallicity fluctuation map as
\begin{equation}
\label{eqn:TPCF}
    \xi(\boldsymbol{r}) = \frac{\left\langle Z'(\boldsymbol{r} + \boldsymbol{r}') Z'(\boldsymbol{r}') \right\rangle}{\left\langle Z'(\boldsymbol{r}')^2\right\rangle},
\end{equation}
where $Z'(\boldsymbol{r})$ is the metallicity fluctuation at position $\boldsymbol{x}$ in the map, and the angle brackets $\left\langle\cdot\right\rangle$ denote averaging over the dummy position variable $\boldsymbol{r}'$. In practice we are interested only in the two-point correlation as a function of scalar separation $r$ rather than vector separation $\boldsymbol{r}$. We compute this by taking every pair of pixels $(i,j)$ in the deprojected, mean-subtracted map, computing their separation $r_{ij}$, and binning the pixel pairs by $r_{ij}$. We then compute the two-point correlation function for the $n$th bin, spanning the range of separations $(r_n, r_{n+1})$ as
\begin{equation}
    \xi_n = \frac{\sigma_{Z'}^{-2}}{N_n} \sum_{r_n < r_{ij} \leq r_{n+1}} Z'_i Z'_j,
    \label{eq:xi_definition}
\end{equation}
where the sum runs over the $N_n$ pixel pairs $(i,j)$ for which $r_n < r_{ij} \leq r_{n+1}$, and
\begin{equation}
    \sigma_{Z'}^2 = \frac{1}{N_p} \sum_{i=1}^{N_p} {Z'_i}^2
\end{equation}
is the total variance of the $N_p$ pixels in the map.

We estimate the uncertainty of the two-point correlation function using a Monte Carlo approach. For each pixel $i$ in the original map, we have a line flux $f_i$ and an uncertainty of the line flux $\sigma_{f, i}$. In each bootstrap trial, we generate a random line flux map by drawing a value for each pixel $i$ from the distribution $N(f_i, \sigma_{f, i})$, where $N(\mu, \sigma)$ denotes a normal distribution with mean $\mu$ and standard deviation $\sigma$. We repeat this process for every line in our metallicity diagnostic (i.e. H\textsc{$\alpha$}, [N \textsc{ii}]$\lambda6584$, [S \textsc{ii}]$\lambda6717$, and [S \textsc{ii}]$\lambda6731$). We then repeat the steps of our analysis pipeline, including adaptive binning and reconstruction of the full line flux and metallicity field, to derive a realisation of the two-point correlation function. We repeat this process 50 times, thereby deriving a sample of 50 two-point correlations. We take the mean and standard deviation of these 50 realisations as our final two-point correlation function $\xi_{\rm obs}$ and its uncertainty $\sigma_{\xi, \rm obs}$.

We finally carry out a parametric fit to our derived two-point correlation function using the functional form proposed by \citetalias{KT18}. The model is based on two key processes, the production of the metals and their redistribution. We modify the model to account for the effects of beam smearing and noise as described in Section 3.3 and Appendix D of \citetalias{L21}. For convenience, we repeat here the functional form to which we fit:
\begin{eqnarray}
\lefteqn{\xi_{\rm model}(r) = \frac{2}{\ln\left(1 + \frac{2\kappa t_*}{\sigma_0^2/2}\right)
} \left[\frac{\Theta(r-\ell_{\rm pix})}{f} + \Theta(\ell_{\rm pix}-r)\right]
} \qquad\qquad
\nonumber \\
& &
\int_0^\infty e^{-\sigma_0^2 a^2/2} \left(1 - e^{-2 \kappa t_* a^2}\right) \frac{J_0(ar)}{a} \, da,
\label{eqn:model}
\end{eqnarray}
where $\sigma_0^2 = \sigma_{\rm beam}^2 + 2 w_{\rm inj}^2$, $\sigma_{\rm beam}$ is the dispersion (not the FWHM) of the observational beam in physical distance, $w_{\rm inj}$ is the physical injection width over which supernovae inject metals, $\kappa$ is the diffusion coefficient, $t_*$ is the star formation duration, $\ell_{\rm pix}$ is the size of a pixel in the observed map, and $f$ is the factor by which observational errors in the derived metallicities increase the variance in the metallicity fluctuations compared to the true variance. Here $\Theta(x)$ is the Heaviside step function, and the purpose of the term in square brackets containing $\Theta(x)$ terms is to account for the fact that errors are perfectly correlated within a single pixel of the observed image, and completely uncorrelated (at least in our approximation) between different pixels. In order to eliminate edge effects in the correlation functions \citep[e.g.][]{Menon21}, we only fit $\xi_{\rm model}(r)$ over a range in $r$ up to 70\% of the median radius (measured relative to the galactic centre) of the available pixels.

\begin{figure*}
\includegraphics[width=1.0\linewidth]{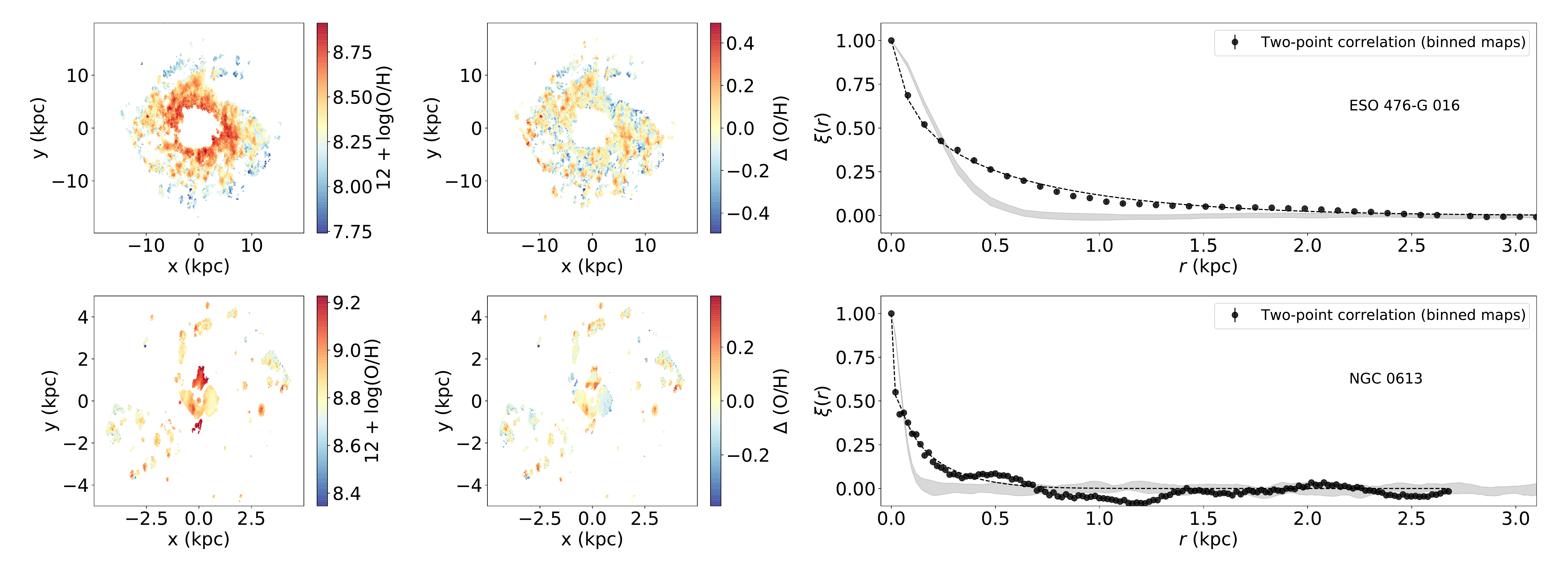}
\caption{Examples of galaxies with different spatial coverage and $L^1$ norm. The upper row shows a galaxy with a good fit (ESO 476-G 016, the host galaxy of ASASSN-18oa, $L^1 = 0.011$), while the lower row shows a poor fit (NGC 0613, $L^1 = 0.030$). From left to right, each row shows the deprojected, adaptively binned metallicity map, the metallicity fluctuation map (i.e., the metallicity map after subtracting the radial gradient), and the two-point correlation function. In the right panels, red dots show the measured two-point correlation function (truncated at the radius where we reach 70\% of the median pixel radius -- see main text), the dashed line shows the best-fitting \citetalias{KT18} model, and the grey band shows the two-point correlation of a beam-smeared noise map with the same masking as the galaxy map.}
\label{fig:l1_norm_illus}
\end{figure*}

We use the measured two-point correlation function $\xi_{\rm obs}$ and its uncertainty $\sigma_{\xi, \rm obs}$ for each galaxy in our sample to generate posterior distributions for the parameters of \autoref{eqn:model} using the \textsc{python} package \textsc{emcee} \citep{emcee}, an implementation of an affine-invariant ensemble sampler for Markov chain Monte Carlo (MCMC). We focus here on four parameters: $\sigma_{\rm beam}$, $w_{\rm inj}$, $\kappa t_*$, and $f$. The likelihood function is given as
\begin{eqnarray}
\lefteqn{\ln p(\xi \mid \sigma_{\rm beam}, w_{\rm inj}, \kappa t_*, f) =
}
\nonumber \\
& &
-\frac{1}{2} \sum_n \left[\frac{(\xi_{\rm model} - \xi_{\rm obs})^2}{\sigma_{\xi, \rm obs}^2} + \ln(\sigma_{\xi, \rm obs}^2)  \right],
\end{eqnarray}
where quantities subscripted `model' are evaluated from \autoref{eqn:model}, quantities subscripted `obs' refer to the correlation functions measured in the observations, and the sum is over the $n$ bins for which we have evaluated the observed correlation function. We adopt flat, uninformative priors on $w_{\rm inj}$, $\kappa t_*$, and $f$ for all positive values (and zero for negative values). We adopt an informative prior for $\sigma_{\rm beam}$, which follows a Gaussian functional form $N(\sigma_{\rm beam,0}, \sigma_{\rm beam,std})$, where the central estimate $\sigma_{\rm beam,0}$ and standard deviation $\sigma_{\rm beam,std}$ are computed by translating our central estimate $\mathrm{FWHM}_0$ and uncertainty $\mathrm{FWHM}_{\rm std}$ on the beam FWHM into physical dispersions in the beam size as projected onto the galactic disc. The FWHM central estimate and uncertainty are computed via the procedure described in \aref{app:psf}. This approach allows us to marginalise the remaining parameters over our uncertainty in the exact size of the PSF.

Our adaptive binning procedure introduces one extra complication in our fitting procedure that did not arise in \citetalias{L21}. While this procedure allows us to obtain much better estimates of the correlation length, $\sqrt{\kappa t_*}$, than would be possible using the unbinned maps that have large gaps where the S/N is too low to allow reliable metallicity measurements, it also potentially produces bias in our estimates of $w_{\rm inj}$. This is because in our model (\autoref{eqn:model}), the correlation function is only sensitive to a quadrature sum of the injection and beam widths, $\sigma_0^2 = \sigma_{\rm beam}^2 + 2 w_{\rm inj}^2$. Thus while our posterior on $\kappa t_*$ is insensitive to the beam width (as confirmed numerically in \citetalias{L21}), the injection width $w_{\rm inj}$ is highly sensitive to it. A side effect of binning is that the effective beam width, $\sigma_{\rm beam}$, is not uniform across the map. As we show in \aref{app:binning}, this makes it problematic to derive a posterior distribution for $w_{\rm inj}$, since then there is no unique beam width.

For this reason, we carry out our MCMC estimation of the posterior probability density functions (PDFs) in two stages. In the first stage, we use $\xi_{\rm obs}$ and $\sigma_{\xi, \rm obs}$ derived from the adaptive-binning-reconstructed map using the priors as described above. In the second stage, where we expect to be able to obtain reliable estimates of $w_{\rm inj}$, we fit to correlation functions computed from the original, non-binned maps. In this stage all the priors are the same as above, except that rather than a flat prior for $\kappa t_*$, we use as our prior a Gaussian kernel density estimate (KDE) of the \textit{posterior PDF} of $\kappa t_*$ derived in the first stage. This approach allows us to compute an estimate for $w_{\rm inj}$ using the unbinned map for which $\sigma_{\rm beam}$ is well-defined, but with $\kappa t_*$ constrained by the more accurate estimates derived from the binned map. In the results we report below, posteriors for $\kappa t_*$ are always those derived from the first stage (using the filled, adaptively-binned maps), while posteriors for $w_{\rm inj}$ are those derived from the second stage (using the non-binned, sparser maps).

We carry out each stage of the MCMC fit using 100 walkers, run for 1000 steps in total; visual examination shows that the chains are well-converged after $\sim 500$ steps, so we discard the first 500 steps for burn-in, and derive the posterior PDF from the final 500 steps.

\subsection{Robustness of the method and sample down-selection}
\label{subsec:selection}

The AMUSING++ compilation has many galaxies in the sample that are only sparsely filled by star-forming regions producing enough ionising radiation to allow us to measure the metallicity; adaptive binning improves but does not completely remedy this situation. Moreover, because many of the AMUSING++ galaxies were originally observed for reasons other than to study the galaxy itself (e.g., many were observed to analyse the host environment of supernovae), the MUSE FoV is in some cases not well aligned with the galaxy. We must therefore test how well our pipeline performs under such conditions. In \citetalias{L21} we showed that our measurements of the correlation function from CALIFA required a minimum number of pixels in order to be robust. This approach worked well for the diameter-selected CALIFA sample, which was designed such that the targeted galaxies filled the IFU. A new approach is needed for AMUSING++, which lacks this uniformity. In this work we therefore take a different approach that assesses both the number of pixels available for analysis and their distribution within the MUSE FoV. 

Formally, for each galaxy we define two statistics: the number filling factor and the area filling factor using the highest S/N maps from H$\alpha$ emission. We define the number filling factor as the ratio of the number of pixels for which the S/N ratio in all lines is $>3$ (i.e., those for which we can measure a metallicity) to the number for which the H\textsc{$\alpha$} line is detected at S/N $>3$ in the binned map; we take the number of H$\alpha$-detected pixels to describe the approximate size of the full galaxy. We define the area filling factor as the ratio of the area covered by pixels where all lines are detected at S/N $>3$ to the area of the convex hull defined by the set of pixels where H$\alpha$ is detected at S/N above 3. Compared with the simple number filling factor, the area filling factor reflects the pixel distribution. If pixels for which we can derive the metallicity are sparse, even though the number filling factor is high, the area filling factor will be relatively low because there will be many uncovered pixels in the convex hull. The number filling factor will always be greater than or equal to the area filling factor, and with equality holding when the H$\alpha$-detected pixels form a convex set.

\begin{figure}
\includegraphics[width=1.0\linewidth]{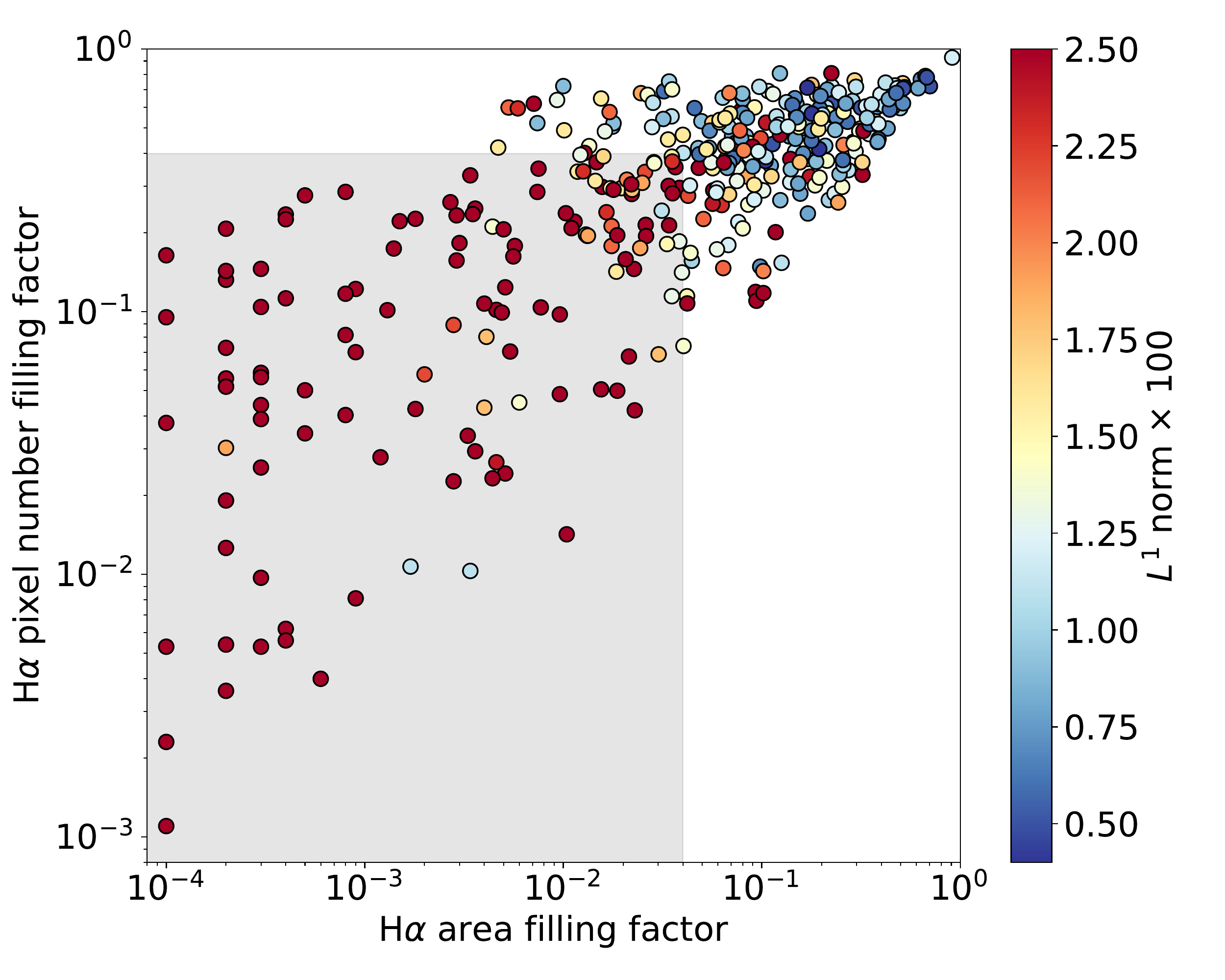}
\caption{The distributions of area ($x$ axis) and pixel number ($y$ axis) filling factors, colour coded by $L^1$ norm. The shaded area denotes the region where we reject the samples. Most of the high $L^1$ norm circles are located in the shaded area, which is as expected since these galaxies discs are too sparsely sampled to allow reliable parameter estimates. We also reject the red circles ($L^1\geq 0.025$) outside the shaded area, but these are only a small minority of the sample.}
\label{fig:fill_frac}
\end{figure}

\begin{table*}
\caption{Global properties, correlation lengths, and injection widths of AMUSING++ galaxies. Columns are as follows: (1) AMUSING++ name; (2) position angle; (3) $b/a$ axis ratio; (4) distance; (5) $r$-band effective radius; (6) FWHM of PSF; (7) stellar mass; (8) H$\alpha$ star formation rate; (9) - (10) correlation length and injection width derived in this work, respectively. For $l_{\rm corr}$ and $w_{\rm inj}$, the central value is the 50th percentile of the posterior PDF, and the error bars show the 16th to 84th percentile range. For cases where one of our fit parameters is not well-constrained (see \autoref{sec:results}) we only report the 86th percentile value as an upper limit (e.g. $w_{\rm inj}$ of NGC 1483). Columns (2) - (8) are from \citet{Lopez-Coba20} and S\'anchez et al. (2022, in preparation). The $R_e$ values are $r$-band half-light radii derived from an isophotal analysis. The SFR values are derived from dust-corrected H$\alpha$ \citep{Sanchez21}. This table is a stub to show structure; the full table is available in the electronic publication.}
\begin{tabular}{cccccccccc}
\hline
Name & PA & $b/a$ & $D$ & $R_e$ & PSF & $\log(M_*/\mathrm{M}_{\odot})$ & $\log(\mathrm{SFR}/\mathrm{M}_{\odot}\mathrm{yr}^{-1})$ & $l_{\rm corr}$ & $w_{\rm inj}$ \\
& ($^{\circ}$) & & (Mpc) & (kpc) & ($\arcsec$) & & & (kpc) & (pc) \\
(1) & (2) & (3) & (4) & (5) & (6) & (7) & (8) & (9) & (10)\\
\hline
NGC 1042 & 17 & 0.445 & 19.0 & 2.96 & 0.798 & $9.879\pm0.093$ & $-1.097\pm0.059$ & $0.241_{-0.002}^{+0.002}$ & $46_{-28}^{+16}$ \\%[1.25ex]
\\
NGC 1068 & 118 & 0.782 & 16.0 & 1.68 & 1.440 & $11.383\pm0.111$ & $0.170\pm0.062$ & $0.321_{-0.002}^{+0.002}$ & $7_{-1}^{+2}$ \\%[1.25ex]
\\
NGC 1483 & 49 & 0.714 & 16.1 & 1.60 & 0.744 & $9.033\pm0.091$ & $-0.696\pm0.059$ & $0.217_{-0.003}^{+0.002}$ & $<6$ \\%[1.25ex]
\hline
\label{tab:full_results}
\end{tabular}
\end{table*}

\begin{figure*}
\includegraphics[width=1.0\linewidth]{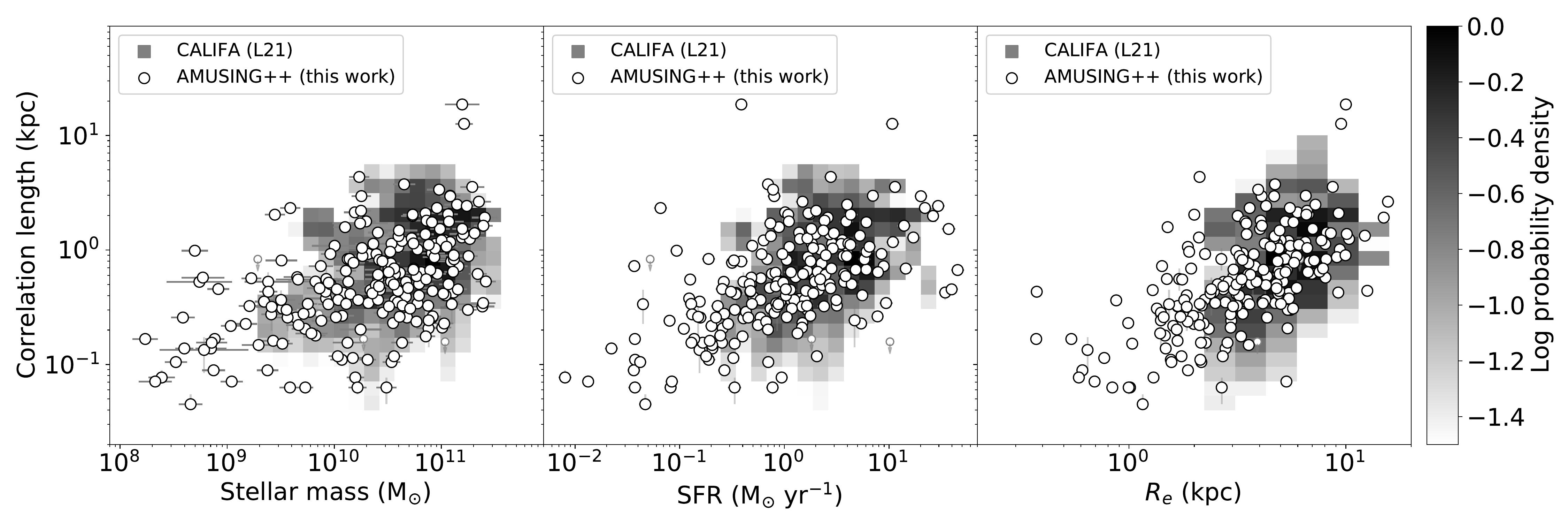}
\caption{Correlation length versus stellar mass (left), SFR (middle), and $R_e$ (right). In this figure, large circles with error bars show the 50th percentile results from this study, with the error bars showing the 16th to 84th percentile range. Small circles with downward arrows indicate 86th percentile upper limits, as indicated in the text. For comparison the background heat map shows the results of \citetalias{L21}; probabilities are normalized so that the maximum pixel value in each panel is unity. The Pearson correlations for the distributions shown in each panel (computed using data from the present study only) are 0.50 (stellar mass), 0.53 (SFR), 0.62 ($R_e$).}
\label{fig:prop_lcorr}
\end{figure*}

\begin{figure}
\includegraphics[width=1.0\linewidth]{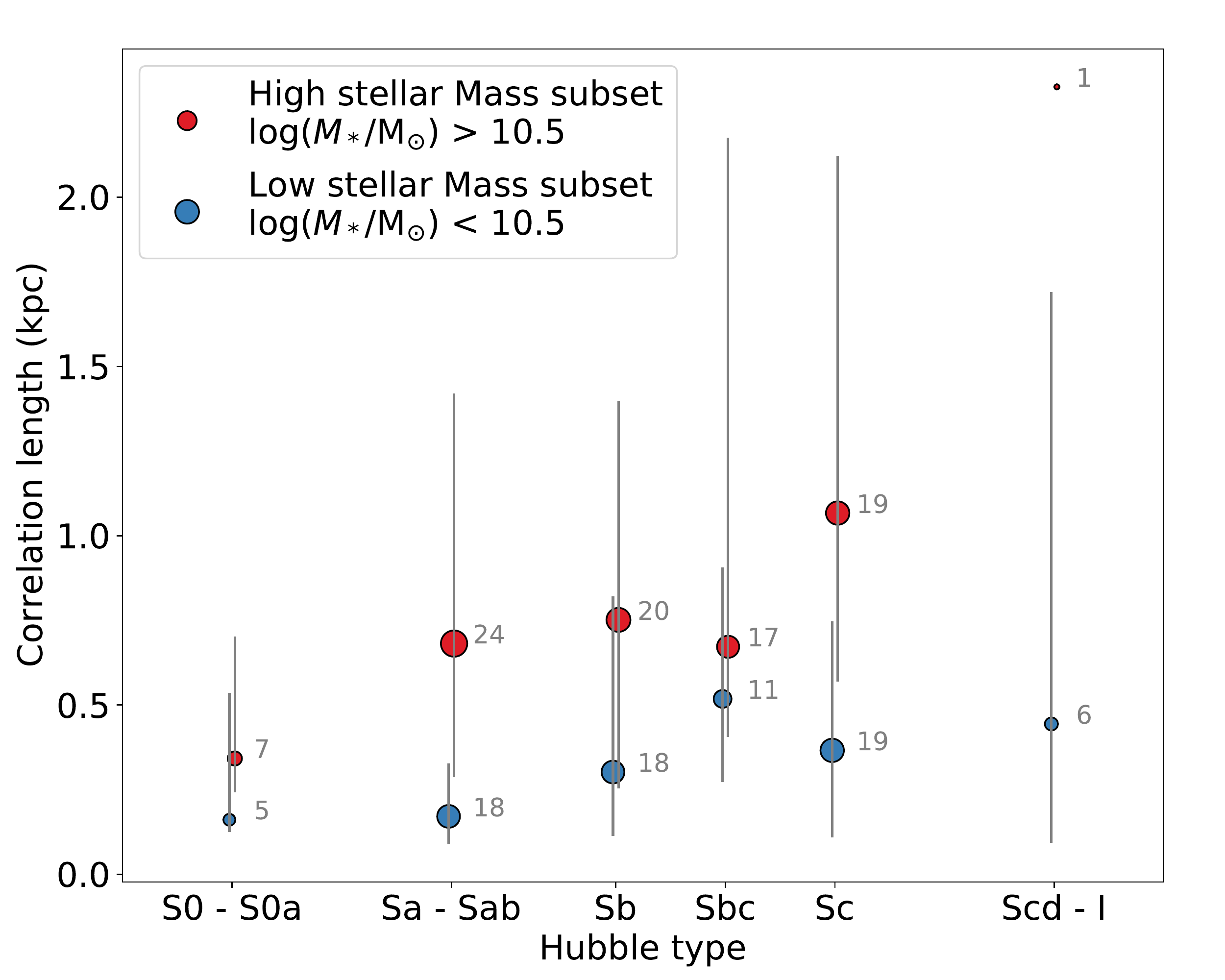}
\caption{Median correlation lengths of galaxies grouped by Hubble type and divided into two subsets, low stellar mass [$\log(M_*/$M$_{\odot})<10.5$, blue circles] and high stellar mass [$\log(M_*/$M$_{\odot})>10.5$, red circles]. The size of each circle shows the number in the subset, which is also annotated next to the marker. The error bar shows the 16th and the 84th percentiles of $l_{\rm corr}$ in each subset.}
\label{fig:morph}
\end{figure}

\begin{figure*}
\includegraphics[width=1.0\linewidth]{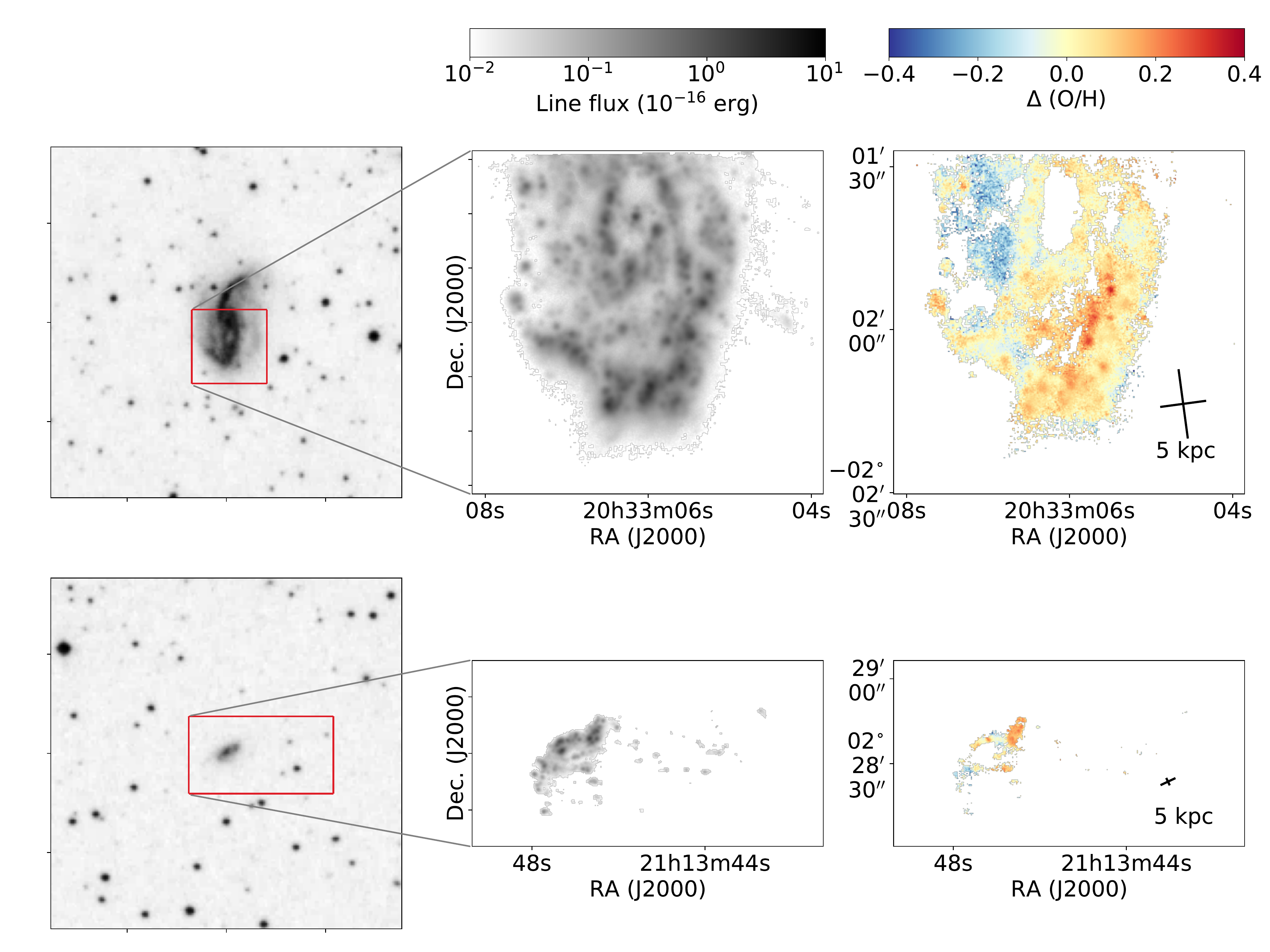}
\caption{Broad band images (the Digitized Sky Survey Palomar Sky Survey I red) of NGC 6926 (upper left) and JO206 (lower left), H$\alpha$ line flux maps (upper middle and lower middle, respectively), and metallicity fluctuation maps (upper right and lower right, respectively). The red boxes illustrate the MUSE observation coverage. Since all of them are before deprojection, the longer bar in a cross shows the scale in the direction of the major axis (position angle), and the shorter bar shows the minor axis.}
\label{fig:large_lcorr}
\end{figure*}

The number and area filling factors defined above allow us to characterise how well the galaxy is sampled, but they do not by themselves tell us what level of sampling is required to generate reliable results. As a proxy for this question, we ask the closely related question of how well we can fit a \citetalias{KT18} model (\autoref{eqn:lcorr_model}) to the data. We characterise the offset between the observed $\xi_{\rm obs}$ and best-fitting model $\xi_{\rm model}$ two-point correlation functions (using the model determined during the first MCMC stage) using the $L^1$ norm defined as
\begin{equation}
\begin{aligned}
\label{eqn:l1_norm}
L^1 & = \frac{1}{R} \int_{0}^{R} | \xi_{\rm model} - \xi_{\rm obs} | \,\mathrm{d}R \\
& = \frac{\Delta R}{R} \sum_{n} | \xi_{\rm model} - \xi_{\rm obs} | \\
& = \left\langle | \xi_{\rm model} - \xi_{\rm obs} | \right\rangle,
\end{aligned}
\end{equation}
where $R$ is the galactocentric radius, and $\Delta R$ is the radial extent of one of the bins over which we compute our two point correlation functions. Intuitively, the $L^1$ norm is simply the mean difference between the observed two-point correlation function and the best-fitting \citetalias{KT18} model. \autoref{fig:l1_norm_illus} shows an example comparison between two galaxies, one with a good (low $L^1$ norm, high filling factor) fit and one with a bad (high $L^1$ norm, low filling factor) fit. It is clear that two-point correlation function in the ``good fit'' example is a smooth function of separation and is well-described by the \citetalias{KT18} model, while the ``poor fit'' example, which has a much sparser map, has significant non-monotonic structure in its correlation function. It is possible that this represents not real structure but an artefact of the very sparse spatial coverage, but in either case this galaxy is not well-described by a \citetalias{KT18} model.

\autoref{fig:fill_frac} illustrates the distribution of $L^1$ norm in the space of area and number filling factor. There is a clear correlation between filling factors and $L^1$ norm: the majority (roughly 82\%) of high $L^1$ norm values ($L^1 >0.025$) lie in the lower left shaded area, while, conversely, the vast majority of the galaxies outside this area are well-fit by a \citetalias{KT18} model. We therefore remove galaxies from our sample if they fall within this region, corresponding to an area filling factor less than 0.04 and a number filling factor less and 0.4. This selection removes 210 of the galaxies in the sample. We also choose to remove 18 galaxies with $L^1\geq 0.025$ outside the shaded area from the sample; while these galaxies do not suffer from poor sampling, they are also not well-described by the model we fit to them and we therefore wish to avoid drawing conclusions based on the outcomes of those fits.  The above selections reduce our initial sample of 447 galaxies to a final sample of 219 galaxies.

\section{Results}
\label{sec:results}

The primary output of our analysis pipeline is a set of posterior PDFs for the two dimensional parameters that characterise our parametric model: the injection width $w_{\rm inj}$ that characterises the size of the region into which metals are first injected by supernovae (SNe), and the correlation length
\begin{equation}
l_{\rm corr} \equiv \sqrt{\kappa t_*}
\label{eqn:lcorr_def}
\end{equation}
that characterises the strength of the mixing in the ISM that occurs after the metals are injected. We report the full set of results in \autoref{tab:full_results}.

For each of these quantities, we derive the median and 68\% confidence intervals. One subtlety that arises in computing confidence intervals is that, for some cases (particularly for $w_{\rm inj}$), at the available resolution of the observations our posterior PDFs are essentially flat all the way to zero, because the signatures of $w_{\rm inj}$ or $l_{\rm corr}$ in the PDF are hidden within the part of the two-point correlation function where the shape is determined purely by the beam size $\sigma_{\rm beam}$. In this situation we only obtain a meaningful upper limit, and while we could still formally compute confidence intervals from our MCMC results, their lower ends would be solely determined by our choice of prior. While it is possible to detect this situation by visual inspection of the posterior PDF, given the size of the sample it is preferable to have an automated criterion. To construct such a criterion, we note that our priors on $w_{\rm inj}$ and $l_{\rm corr}$ are flat at values $>0$, so in the situation we have just described where the posterior mirrors this shape, we should find that the ratio $p_n/p_m = n/m$, where $p_n$ and $p_m$ are the $n$th and $m$th percentiles of the marginal posterior PDF. That is, if the posterior is flat from 0 to the 50th percentile value, then the value of the 5th percentile is simply 1/10 of the value of the 50th percentile. By contrast, if the posterior is peaked rather than flat, we would expect $p_5/p_{50} > 1/10$, with $p_5/p_{50} \to 1$ as our posterior approaches an infinitely sharp $\delta$-distribution. Our automated check is therefore to evaluate $p_{10}/p_{50}$ and $p_{16}/p_{50}$. If we find that either $p_{10}/p_{50} \leq 1/5$ or $p_{16}/p_{50} \leq 1/3$, we report the result as an upper limit at the 86th percentile value, rather than giving a full confidence interval. Similarly, we plot such points as upper limits using arrows pointing downwards in the figures below.

\subsection{Correlation length}
\label{subsec:lcorr}

In \autoref{fig:prop_lcorr} we report correlation length as a function of $M_*$ (left, \textsc{pipe3d} fitted), SFR (middle, dust-corrected H$\alpha$), and $R_e$ (right, isophotal $r$ band). The previous results from the CALIFA survey ($\sim 800$ pc resolution, \citetalias{L21}) are also illustrated using heat density maps. The Pearson correlations for the distributions from this work shown in \autoref{fig:prop_lcorr} are 0.50 ($M_*$), 0.53 (SFR), 0.62 ($R_e$). To estimate the uncertainties on the correlation coefficients, we draw samples of $l_{\rm corr}$ from the posterior PDF for each galaxy, compute the Pearson correlation between $l_{\rm corr}$ and a galaxy property for one realization, and repeat the process 50 times. This gives a typical uncertainty of Pearson correlation less than 0.01. It is clear that are results are generally consistent with those of the lower resolution CALIFA survey (\citetalias{L21}), but that with the current dataset the correlation between $l_{\rm corr}$ and other galaxies is both tighter and extends to significantly lower mass, SFR, and $R_e$. To first order the relationship is between $l_{\rm corr}$ and the other galaxy quantities is roughly consistent with a power law with a slope of $1/2$ for stellar mass and SFR, i.e., roughly $l_{\rm corr} \propto M_*^{1/2} \propto \mathrm{SFR}^{1/2}$, and 1 for $R_e$, i.e., roughly $l_{\rm corr} \propto R_e$. There is marginal evidence for a departure from this scaling at low SFR and $R_e$, with correlation lengths not falling below a few hundred pc regardless of galaxy size. As in \citetalias{L21}, since stellar mass, SFR, and $R_e$ are all correlated with one another in our galaxy sample, it is difficult to determine which of these quantities plays a primary role in determining $l_{\rm corr}$, and which are secondary correlations. We present a possible interpretation of the results in light of a simple theoretical model below.

In addition to these correlations with galaxy size and mass, several authors have posited that large-scale structures in galaxy discs (e.g. spiral arms or bars) are responsible for ISM chemical mixing \citep[e.g.][]{DiMatteo13, Grand16, Sanchez-Menguiano17, Spitoni19, Sanchez-Menguiano20_arm}. AMUSING++ does not provide a quantitative estimate of bar strength, but it does provide a Hubble type, which we can use as a rough proxy for the strength of spiral arm structure in our sample. In \autoref{fig:morph} we show median correlation lengths of galaxies grouped by Hubble type, from S0 to I. We also divide the whole sample into high and low stellar mass groups based on the median value $\log(M_*/\mathrm{M_{\odot}}) = 10.5$ of the sample; these two groups are shown in red and blue colours. The size of each circle denotes the size of the samples in each subset, and the error bar shows the standard error. In each Hubble type, the high $M_*$ branch is always higher than the low $M_*$ branch, consistent with the results in \autoref{fig:prop_lcorr}. From Sa to Sc there is a weak but clear trend toward increasing $l_{\rm corr}$ in the high-mass subset of the sample, even ignoring the final Scd-I bin that contains only one galaxy. This partly agrees with the findings of \citetalias{L21}, and suggests that increasing spiral arm strength slightly increases the correlation length, e.g., with S0 galaxies with no spiral arms showing lower average correlation lengths compared to Sc galaxies. However the effect is on order $\lesssim 1\sigma$, and there is no obvious trend in the low-mass sub-sample.

\begin{figure}
\includegraphics[width=1.0\linewidth]{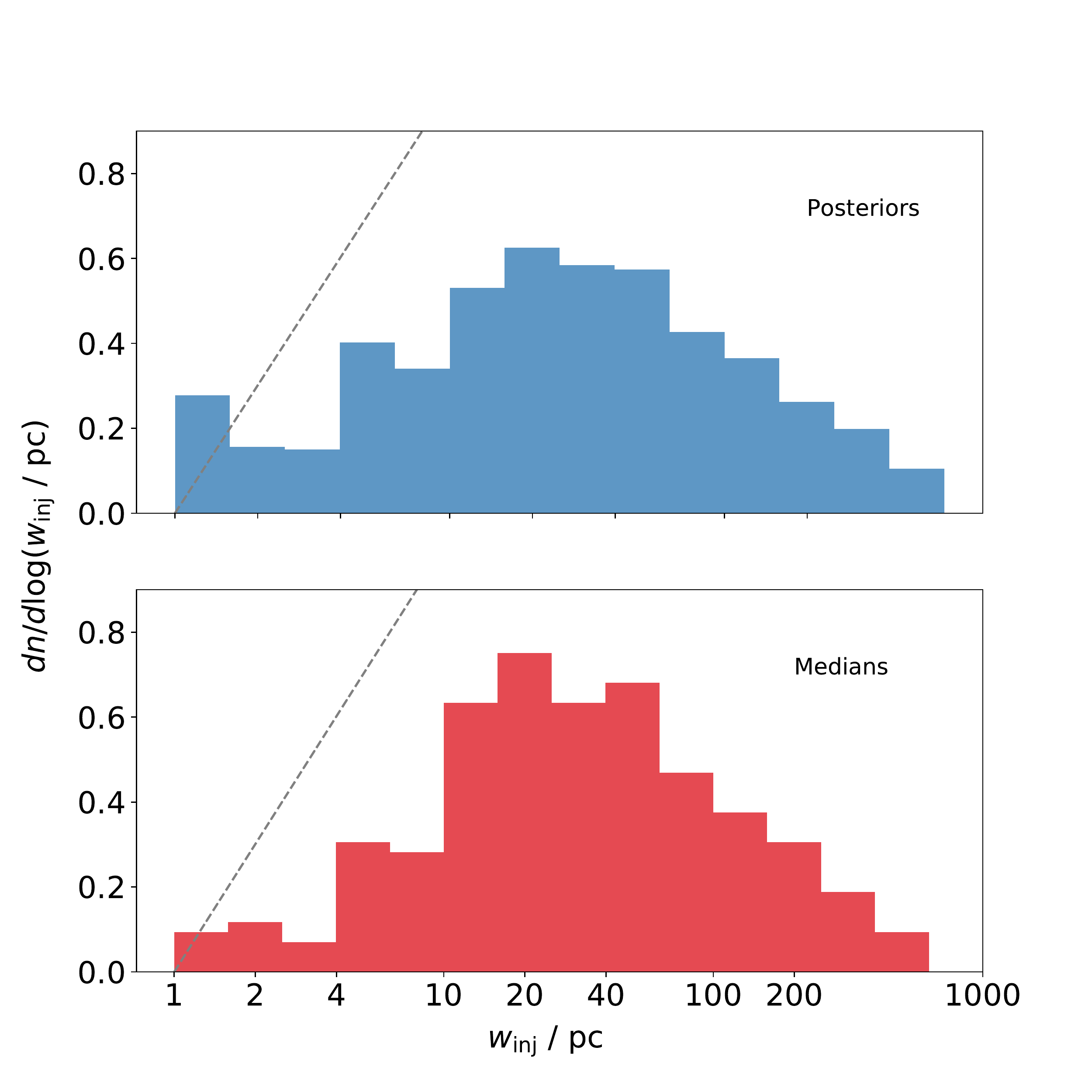}
\caption{Posterior probability density functions for the injection width $w_{\rm inj}$. In the upper panel, the distribution shown is the average of the posterior PDFs derived for all galaxies, weighting each galaxy equally, while in the lower panel we show the distribution of the median values of $w_{\rm inj}$ derived for every galaxy. The grey dashed lines in each panel show the shape of our prior on $w_{\rm inj}$ for comparison.}
\label{fig:winj_hist}
\end{figure}

\begin{figure*}
\includegraphics[width=1.0\linewidth]{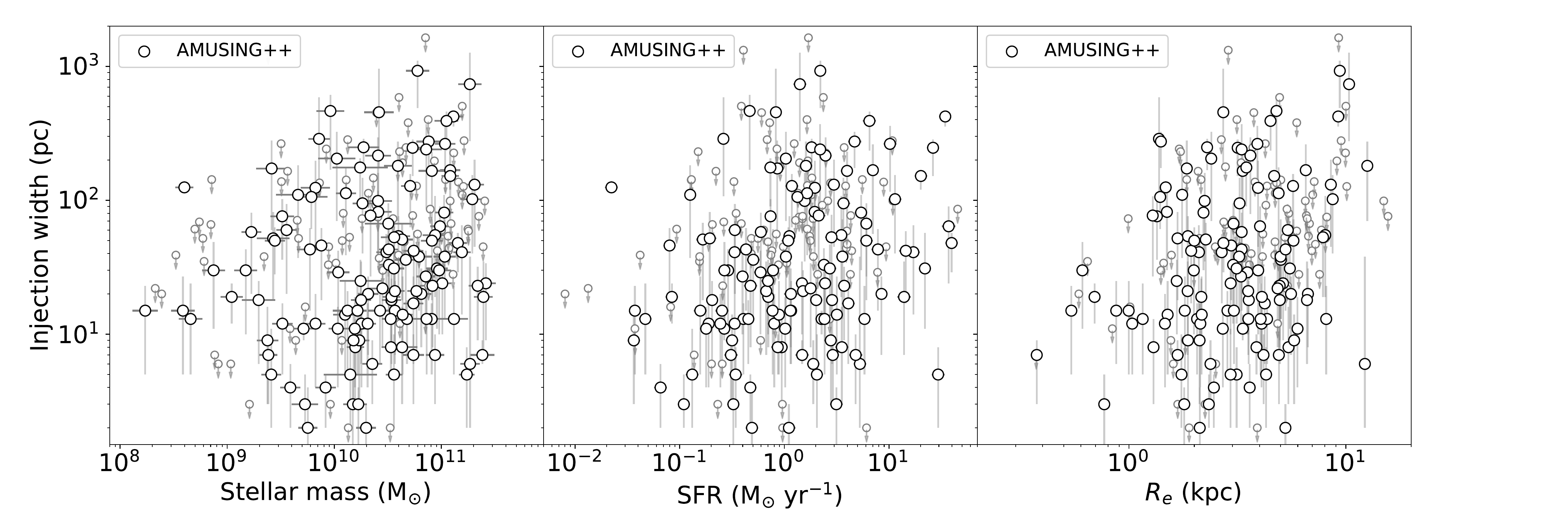}
\caption{Injection width versus stellar mass (left), SFR (middle), and $R_e$ (right). The Pearson correlations for the distributions shown in each panel are 0.18 (stellar mass), 0.17 (SFR), 0.28 ($R_e$). As in \autoref{fig:prop_lcorr}, large points mark 50th percentile values, with vertical error bars showing the 16th to 84th percentile range; small point show 86th percentile upper limits for galaxies were we only derive an upper limit.}
\label{fig:prop_winj}
\end{figure*}

\subsection{Extremely large correlation lengths in interacting galaxies}

We find two galaxies that have extremely large correlation lengths $l_{\rm corr} > 10$ kpc (cf.~\autoref{fig:prop_lcorr}), which are outliers with respect to the rest of the sample that has $l_{\rm corr} < 5$ kpc. They are NGC 6926 and JO206, and we show their (undeprojected) H$\alpha$ maps and metallicity fluctuation maps in \autoref{fig:large_lcorr}. Both galaxies show large non-axisymmetric metallicity fluctuations, demonstrating that the pipeline two-point correlation function computation is detecting a real feature of the galaxy metallicity distributions.

NGC 6926 is a spiral luminous infrared galaxy (LIRG) with a Seyfert nucleus, and is at a very early stage of interaction with the dwarf galaxy NGC 6929 that lies 4 arcmin to its east \citep[outside the MUSE FoV;][]{Dopita15, Herrero-Illana17}. JO206 is a jellyfish galaxy experiencing gas stripping caused by ram-pressure during its infall into a cluster \citep{Ramatsoku19}. These interactions provide a natural explanation for why these two particular galaxies should have metallicity maps that differ so much from the rest of the sample; they also corroborate the trend noted in \citetalias{L21} that interacting galaxies have larger correlation lengths. Since the \citetalias{KT18} model does not take such complicated circumstances into account, e.g. merging / interactions and ram-pressure stripping, there is no reason to expect the simple prediction shown in \autoref{eqn:lcorr_def} to describe the two-point correlation functions for such galaxies. That we are nonetheless able to achieve a reasonable fit should serve as a caution against over-interpreting the goodness of fit for the rest of the sample; the fact that the two-point correlation function is close to the functional form predicted by \citetalias{KT18} does not, by itself, confirm that this very simple model is accurately describing the physical processes giving rise to that correlation function. On the other hand, the fact that the pipeline returns such a large $l_{\rm corr}$ in itself acts as a warning that we may be looking at a system that is outside the range of galaxies described by the \citetalias{KT18} model.

Given that NGC 6926 is experiencing an interaction, there are two plausible explanations for its unusual metallicity fluctuation map. First, the regions where metallicities are enhanced [$\Delta(\rm O/H) > 0$] roughly match the spiral arms. This could indicate that a burst of the extreme star formation in the arms, triggered by shocks caused by the tidal effect, has led to a real jump in the metallicity, which would then be diffused away as the galaxy settles. Second, it is also possible that the map of NGC 6926 simply reflects the failure of our metallicity diagnostic. The D16 diagnostic is calibrated based on H~\textsc{ii} regions in the Milky Way and similar nearby galaxies, and the H~\textsc{ii} regions found in a starburst like NGC 6926 may explore ranges of dust extinction or ionisation parameter beyond those where the model is reliable. Yet a third possibility, which is likely the explanation for JO206, is that the large-scale non-axisymmetric metallicity structure to which our correlation length is responding is a result of ongoing gas exchange between the disc and gas outside the disc. In the case of JO206, the disc might be blended with the metal-poor gas in circumgalactic medium (CGM), while in NGC 6926 there may be gas exchange between the more metal-rich disc gas and the metal-poor gas in the dwarf with which the disc is starting to merge. These interactions could easily produce real metallicity structures that are correlated on $>10$ kpc scales.

\subsection{Injection width}
\label{subsec:winj}

Thanks to the high spatial resolution of MUSE IFS, our dataset allows the first ever measurements of the injection width $w_{\rm inj}$, which is theoretically expected to be on order $\sim 70$ pc. In the context of the \citetalias{KT18} model, the injection width is the size of the region into which SNe deposit metals directly, without them needing to be carried there by diffusion. Independent of this underlying theoretical picture, the observational manifestation of $w_{\rm inj}$ is a change in the shape of the metallicity correlation function associated with a break in behaviour at intermediate distances: at separations $r \gg w_{\rm inj}$ this shape is determined by the interacting, diffusing metal fields from many sources, while for $r \ll w_{\rm inj}$ the correlation function mostly reflects the correlation of sources with themselves. 

In the upper panel of \autoref{fig:winj_hist} we report the posterior PDF of $w_{\rm inj}$ summed over all galaxies, which each galaxy weighted equally. That is, to construct this plot we have simply made a histogram of the MCMC walker values of $w_{\rm inj}$ (discarding the burn-in phase) over all galaxies, so a single galaxy can contribute to more than one bin, with its contribution to each bin proportional to the integral of the PDF over that bin. In the lower panel of \autoref{fig:winj_hist}, by contrast, we show the histogram of median values of $w_{\rm inj}$ for the sample. In both panels we add a grey dashed line illustrating the flat prior (in linear space), for the sake of comparison. The fact that our posterior PDFs have a shape nothing like that of the grey dashed line shows that our results are far from a simple reflection of the prior, and instead reflect a real detection of the injection width, not simply upper limits.

As with \autoref{fig:prop_lcorr}, in \autoref{fig:prop_winj} we report injection width as a function of $M_*$, SFR, and $R_e$. The is little obvious correlation between $w_{\rm inj}$ and any of these quantities, as is confirmed by a quantitative evaluation of the Pearson correlations; these are 0.18 for $M_*$, 0.17 for SFR, and 0.28 for $R_e$. The fact that these correlations are non-zero most likely reflects the clear trend in the figure that large values of $w_{\rm inj}$ are absent for galaxies with small $M_*$, SFR, and $R_e$, i.e., the upper left quadrants of \autoref{fig:prop_winj} are empty. However, this most likely just reflects the physical size of the maps: galaxies with $R_e \sim 1$ kpc have metallicity maps of comparable size, and thus we are unlikely to measure $w_{\rm inj}$ larger than a few hundred pc, simply because the entire map is not that large. Thus we conclude that even the mild correlation we do see is likely an artefact of finite sizes of the galaxies, rather than a true detection. 

\section{Discussion}
\label{sec:discussion}

\subsection{Correlation length, SFR, and turbulence}

A range of analysis techniques, including the two-point correlation function and its correlation length (\citetalias{L21} and this work), the 30\% correlation scale \citep{Kreckel20, Williams22}, and the semivariogram and its homogeneity scale \citep{Metha21}, broadly agree that the characteristic scale of metallicity correlations within massive galaxies is $\sim 1$ kpc. This consistency from multiple independent teams, techniques, and resolutions suggests that this finding is robust. It is therefore of great interest to understand why galaxy metal fields are correlated at this scale. We consider a possible explanation from first principles in this subsection.

The diffusion process in the ISM is essentially a random walk. In a random walk, the characteristic distance $X$ that a particular metal atom should travel over a time $\tau$ obeys
\begin{equation}
\langle X^2 \rangle = 6\kappa \tau \propto \ell \langle v^2 \rangle \tau.
\label{eqn:rand_walk}
\end{equation}
For a non-turbulent environment, $\ell$ is the mean free path and $\langle v^2 \rangle$ is the mean square velocity of the ensemble of particles, but since the ISM is turbulent, $\ell$ is described by the scale height $h$, which is approximately the size of the largest turbulent eddies (and is the characteristic size scale of the energy source driving the turbulence), and $\langle v^2 \rangle$ is the bulk velocity dispersion $\sigma_{\rm g}^2$. \cite{Karlsson13} proposed $\kappa \approx h \sigma_{\rm g} / 3$ and thus
\begin{equation}
l_{\rm corr} = \sqrt{\frac{1}{3} h \sigma_{\rm g} \tau}.
\label{eqn:lcorr_model}
\end{equation}

\begin{figure*}
\includegraphics[width=1.0\linewidth]{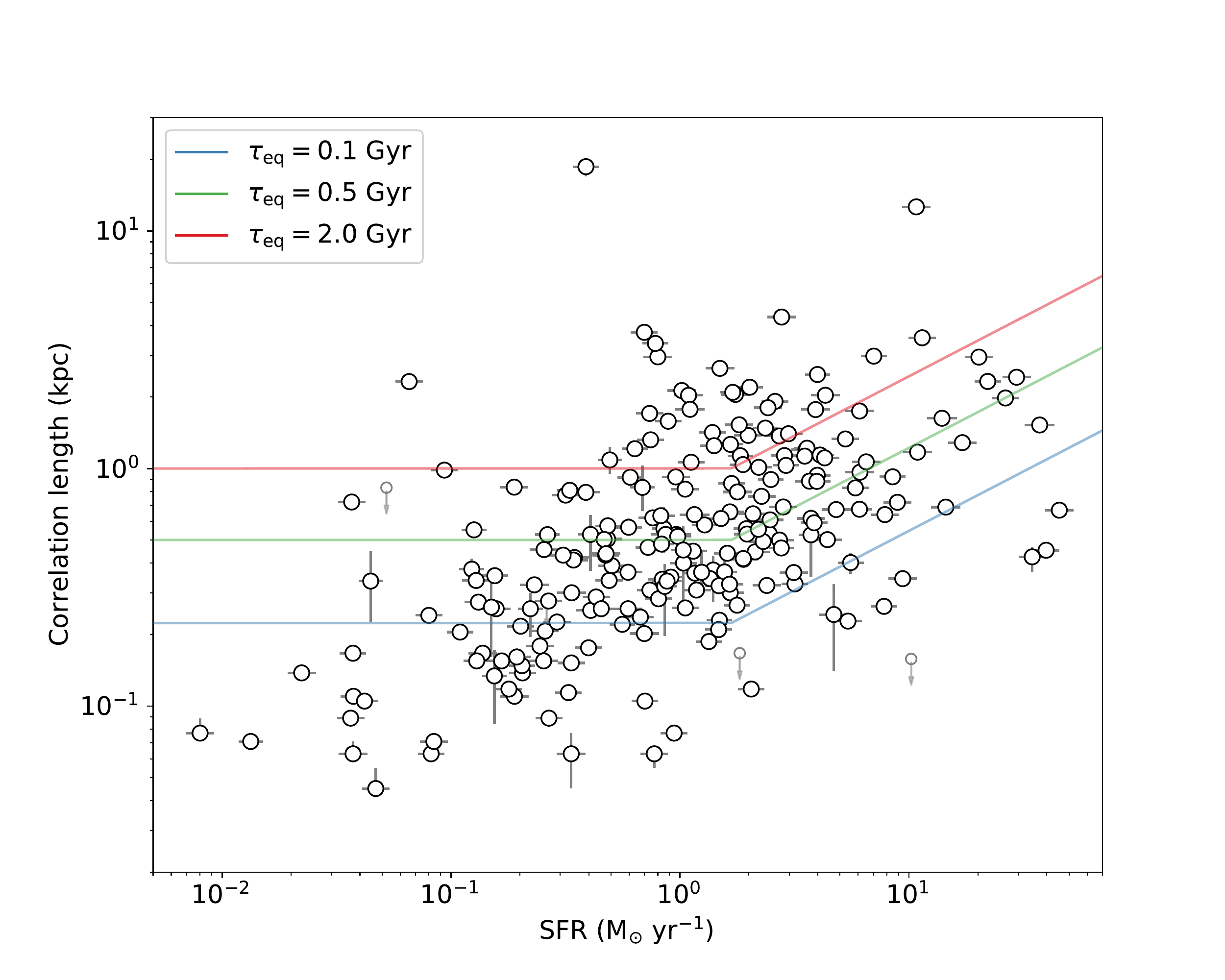}
\caption{Correlation length $l_{\rm corr}$ versus star formation rate (the middle panel of \autoref{fig:prop_lcorr}). Coloured lines show values of $l_{\rm corr}$ predicted using velocity dispersions computed from model of \citep{Krumholz18}. The blue, green, and red lines show results for equilibrium timescales $\tau_{\rm eq} = 0.1, 0.5$, and 2.0 Gyr, respectively.}
\label{fig:lcorr_vs_SFR}
\end{figure*}

It is clear that \autoref{eqn:lcorr_model} has the same mathematical format as \autoref{eqn:lcorr_def} and equation (13) in \citetalias{L21}. However, there is a subtlety in that it is not clear what value one should adopt for the characteristic timescale $\tau$. The \citetalias{KT18} model proposed that $\tau$ should be the entire disc formation timescale $t_*$. However, \cite{Metha21} and \citetalias{L21} find that this choice leads to a predicted correlation length that is too large compared to observations. (\citealt{Metha21} also point out that setting $\tau \approx t_*$ yields an overall variance that is too small). Instead, the data are best fit if $\tau$ is not as long as the age of the galaxy disc and is instead closer to metallicity gradient equilibration timescale $\tau_{\rm grad, eq}$ \citep{Sharda21} or the gas depletion timescale $\tau_{\rm dep} \approx 2$ Gyr \citep[e.g.][]{Bigiel08, Bigiel11, Utomo17}. This discrepancy might be due to the fact that the \citetalias{KT18} model is essentially a closed-box model and does not consider metal-poor gas inflow and the resulting dilution of ISM metallicity, or gas outflow and its effects on ISM metallicity. By contrast, \citet{Sharda21} show that the time $\tau_{\rm grad, eq}$ required for a galaxy to attain a stable metallicity gradient is dictated by the balance between centrally-concentrated star formation steepening the gradient (tempered by loss of some of the metals to galactic winds), and radial inflow, turbulent mixing, and metal dilution caused by gas inflow flattening it; several of these processes are necessarily absent in a closed-box model.

We can make this discussion more quantitative by attempting to estimate the timescale over which we might expect the metal content in a galaxy to be ``reset'' by inflow and outflow, and seeing if this gives a timescale $\tau$ in better agreement with observations. \cite{PM14} propose a model to describe the evolution of galaxies' mean metallicities considering star formation and gas inflow/outflow. They conclude that the time over which the mean metallicity reaches equilibrium $\tau_{\rm mean, eq}$ is less than or equal to $\tau_{\rm dep}$, or more specifically, $\tau_{\rm mean, eq} = \tau_{\rm dep} / (1 - R + \eta)$ where $R\approx 0.4$ is the mass return fraction from stars in the instantaneous recycling approximation \citep{Tinsley80}, and $\eta$ is the mass-loading factor that describes the ratio of the mass outflow rate $\dot{M}$ and SFR. If we replace the star formation timescale $t_*$ in the \citetalias{KT18} model with $\tau \approx \tau_{\rm eq} \lesssim \tau_{\rm dep}$, and adopt typical galactic scale heights $h\approx 150$ pc, velocity dispersions $\sigma_{\rm g} \approx 10$ km s$^{-1}$ \citep[e.g.][]{Leroy08, Walter08, Kalberla09, Levy18}, and timescales $\tau \approx 2.2$ Gyr estimated from $\tau_{\rm dep}$ \citep[e.g.][]{Leroy13}, we find $l_{\rm corr} \approx 1$ kpc. This approach gives a value for metallicity correlation that is in rough agreement with what is measured from local datasets.

To make the comparison more quantitative, we can apply the \citet{Krumholz18} model for the variation of $\sigma_{\rm g}$ with galaxy star formation rate, which agrees well with observations. Using their theoretically predicted the relation between $\sigma_{\rm g}$ and SFR (Equation 60 for the transport+feedback model), while continuing to adopt $h = 150$ pc and $\tau_{\rm eq} \lesssim \tau_{\rm dep} \approx 2$ Gyr, yields a predicted relation between $l_{\rm corr}$ and SFR shown in \autoref{fig:lcorr_vs_SFR}. For reasonable values of $\tau_{\rm eq}$, this prediction is consistent with the data at high SFR. However at low SFR, $l_{\rm corr}$ deviates from the model prediction. One possible explanation is that the considerable mass loss processes that commonly occur in dwarf galaxies may lead to an equilibrium time $\tau_{\rm eq} \ll \tau_{\rm dep}$. The outflow processes leave gas little time to diffuse before being removed from discs. Indeed, several studies report the mass-loading factor $\eta$ is negatively correlated with $M_*$ \citep[e.g.][]{Somerville08, Muratov15, Chisholm17}. Considering that a typical mass loading $\eta$ could be $\sim 10$ at $M_* \approx 10^9$ M$_{\odot}$ and SFR $\approx 0.1$ M$_{\odot}$ yr$^{-1}$, $\tau_{\rm eq}$ could be reduced by a factor $\eta \gg R$ compared to $\tau_{\rm dep}$. However, this reduction would need to be very large to explain the extremely low correlation lengths we find in some dwarfs, since as the figure shows, even $\tau_{\rm eq} = 0.1$ Gyr still leads to correlation lengths that are too large for those found in the lowest SFR galaxies in our sample. This inconsistency could result from either assumptions made here (e.g. scale heights, velocity dispersions) or in the assumed model.

\subsection{Injection width}

We find that typical values of $w_{\rm inj}$ are a few tens of pc (\autoref{fig:winj_hist}), consistent with the observed sizes of typical SN remnants (SNRs), and with the theoretical predictions of \citetalias{KT18}. However, there is a tail of galaxies for which $w_{\rm inj}$ is one to a few hundred pc, more characteristic of the sizes of superbubbles \citep[e.g.][]{Tomisaka81} than of individual SNRs \citep{Draine11}. There is no obvious difference in bulk properties between these galaxies and those for which $w_{\rm inj}$ is smaller, but given the small total number of measurements, this could simply be a result of our sample size being too small to detect correlations.

We find no significant correlation between $w_{\rm inj}$ and $M_*$, SFR, or $R_e$. This is perhaps not unexpected since $w_{\rm inj}$ is a local parameter and our derived $w_{\rm inj}$ should only be an estimated ``overall-averaged'' local fade-away radius of SNe. In the \citetalias{KT18} model $w_{\rm inj}$ should anti-correlate with atomic hydrogen number density. However, we have no direct measurements or indirect proxy for such a parameter in this work and since our sample includes no starbursts, with the possible exception of NGC 6926 ($M_*=1.63\times10^{11}$ M$_{\odot}$, $\mbox{SFR}=10.8$ M$_{\odot}$ yr$^{-1}$), the range of ISM densities spanned by the sample is likely small; this precludes drawing strong conclusions regarding the injection width.

\section{Conclusions}
\label{sec:conclusions}

We apply the two-point correlation function to the maps of galaxy gas-phase metallicities from the AMUSING++ compilation, a sample of 532 galaxies observed with MUSE on the VLT. The combination of instrument and catalogue yields both better spatial resolution and a much larger sample than previous studies of galactic metal distributions. We develop a new technique to enhance the S/N ratio in the data by adaptively binning, in order to compensate for the large areas of galactic discs where weak emission lines (S \textsc{ii}]$\lambda\lambda6717,31$) cannot be detected with acceptable S/N in individual pixels. We then apply the basic analysis pipeline proposed in \citetalias{L21} and show that the metallicity correlations are well described using a simple injection–diffusion model as proposed by \citetalias{KT18}.

One key parameter we derive from the model is the correlation length ($l_{\rm corr}$), which describes the characteristic ISM mixing length scale. We confirm a correlation between $l_{\rm corr}$ and stellar mass, star formation rate (SFR), and effective radius, first identified in \citetalias{L21}, and are able to characterise this correlation with much better statistics thanks to the improved data quality and dataset size available in AMUSING++. We also find find a weak trend of increasing $l_{\rm corr}$ with Hubble type. We show that the trends in $l_{\rm corr}$ with SFR at high SFR can plausibly be explained in terms of the \citet{Krumholz18} transport+feedback model, with predicts an increase in galaxy velocity dispersion with SFR. Extending this explanation to the low-SFR regime requires large mass-loading factors in dwarf galaxies, which are plausible but not independently predicted by the model. We also demonstrate that for some cases when a galaxy is interacting or experiencing gas exchange (e.g. caused by ram pressure), $l_{\rm corr}$ can be extremely large. Inspection of the metallicity in these cases indicates a highly non-axisymmetric distribution, which in turn leads to a substantially larger correlation length. This phenomenon confirms the finding in \citetalias{L21} that merging galaxies show larger $l_{\rm corr}$.

Thanks to the high spatial resolution of MUSE, we are also for the first time able to constrain the injection width ($w_{\rm inj}$), which represents the size of the initial bubble radius of a SN explosion (since oxygen is almost purely enriched by type II SNe). A typical $w_{\rm inj}$ value is $\sim 30$ pc, but the distribution ranges broadly from several pc to over a hundred pc. We find no significant correlation of $w_{\rm inj}$ with any global galaxy property. We note that future instruments will allow further measurements of the metallicity correlations and injection sites for a wider range of galaxies (e.g. MAVIS: \citealt{Rigaut20}; GMTIFS: \citealt{McGregor12}).

In future work we intend to apply this statistical tool to simulated galaxies. This will help address the question of how two-point correlation functions evolve with cosmic time, and whether they are equilibrium or non-equilibrium characteristics of galaxies. We will compare the predicted correlation functions to measured ones, and thereby ask how well simulations are describing the processes of metal transport.

Another potential application of our technique is to compare the two-point correlation functions of nitrogen and oxygen, and in particular to examine their injection widths. Since these elements have different nucleosynthetic sources, they should have substantially different correlation functions, at least at the small scales that are sensitive to the injection width. Nitrogen enhancement is due to deaths of both low-mass stars and SNe, whereas oxygen is purely enriched by the latter. Since nitrogen and oxygen are embedded in the same ISM but have different enrichment sources, nitrogen distributions are expected to possess different two-point correlations with smaller injection widths. Performing this experiment requires access to lines that make it possible to measure N and O abundances independently. The SIGNALS survey \citep{Rousseau-Nepton19} is a good candidate for this purpose thanks to its spatial resolution and wavelength coverage.

\section*{Acknowledgements}

EW \& JTM acknowledge support by the Australian Research Council Centre of Excellence for All Sky Astrophysics in 3 Dimensions (ASTRO 3D), through project number CE170100013. MRK acknowledges support from the Australian Research Council through awards FT180100375 and DP190101258. SFS thanks the support of the PAPIIT-DGAPA IN100519 and IG100622 projects. LG acknowledges financial support from the Spanish Ministerio de Ciencia e Innovaci\'on (MCIN), the Agencia Estatal de Investigaci\'on (AEI) 10.13039/501100011033, and the European Social Fund (ESF) "Investing in your future" under the 2019 Ram\'on y Cajal program RYC2019-027683-I and the PID2020-115253GA-I00 HOSTFLOWS project, from Centro Superior de Investigaciones Cient\'ificas (CSIC) under the PIE project 20215AT016, and the program Unidad de Excelencia Mar\'ia de Maeztu CEX2020-001058-M. ZL thanks Yifei Jin, Henry Zovaro, and Andrew Battisti for generous support and fruitful discussion.

%%%%%%%%%%%%%%%%%%%%%%%%%%%%%%%%%%%%%%%%%%%%%%%%%%
\section*{Data Availability}

The data and the source code underlying this article are available at \href{https://github.com/zidianjun/metallicity-correlation-AMUSING}{https://github.com/zidianjun/metallicity-correlation-AMUSING}.

%%%%%%%%%%%%%%%%%%%% REFERENCES %%%%%%%%%%%%%%%%%%

% The best way to enter references is to use BibTeX:

\bibliographystyle{mnras}
\bibliography{main} % if your bibtex file is called example.bib

% Alternatively you could enter them by hand, like this:
% This method is tedious and prone to error if you have lots of references
%\begin{thebibliography}{99}
%\bibitem[\protect\citeauthoryear{Author}{2012}]{Author2012}
%Author A. N., 2013, Journal of Improbable Astronomy, 1, 1
%\bibitem[\protect\citeauthoryear{Others}{2013}]{Others2013}
%Others S., 2012, Journal of Interesting Stuff, 17, 198
%\end{thebibliography}

%%%%%%%%%%%%%%%%%%%%%%%%%%%%%%%%%%%%%%%%%%%%%%%%%%

%%%%%%%%%%%%%%%%% APPENDICES %%%%%%%%%%%%%%%%%%%%%

\appendix

\section{Image quality of AMUSING++ galaxies}
\label{app:psf}

An accurate estimate of the observational PSF is required for our analysis pipeline to derive an accurate estimate of the metal injection width. Such an estimate is not immediately available for the AMUSING++ compilation, which is assembled from heterogeneous observations made under a range of observing conditions (sometimes including multiple observations of the same field made with very different seeing), and does not provide overall image quality estimate based on fits to point sources in the FoV; indeed, many of the FoVs lack suitable sources.

\begin{figure}
\includegraphics[width=1.0\linewidth]{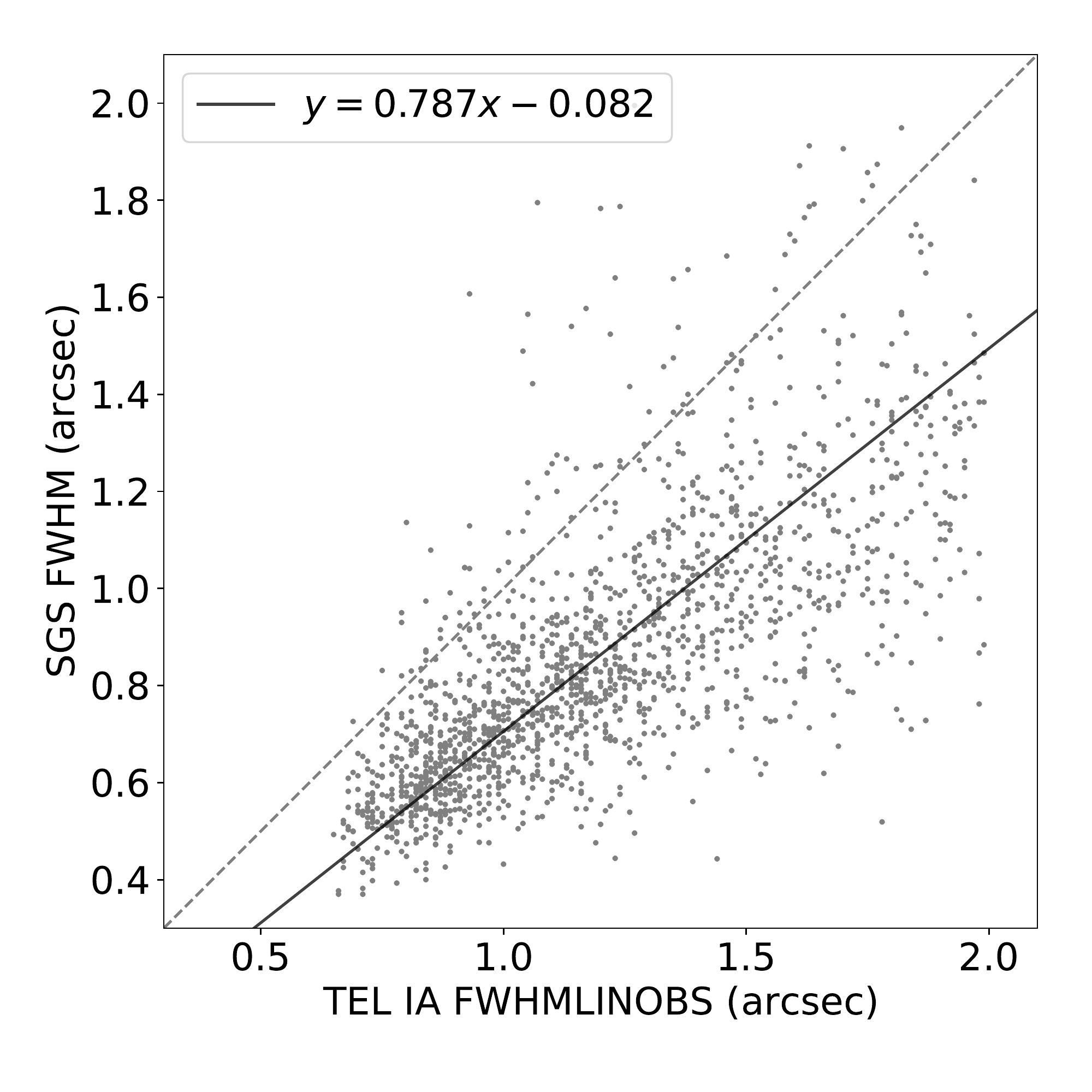}
\caption{SGS FWHM versus TEL IA FWHMLINOBS for each frame in the AMUSING++ compilation. The black solid line is the linear least squares fit to the data, the functional form for which is shown in the legend, while the grey dashed line is the 1-1 line.}
\label{fig:fwhm}
\end{figure}

To estimate the PSF for our pipeline, we start by retrieving the slow-guiding-system (SGS) FWHM and linear fit-delivered seeing FWHM from the telescope image analysis (TEL IA FWHMLINOBS), for every observation of every target in the sample. Of these two, the SGS FWHM is the more accurate estimate of the seeing for MUSE, since this system observes the PSF as the science instrument in essentially the same part of the sky. In contrast, TEL IA FWHMLINOBS reflects measurements made by the active optics system, and thus derives from the PSF seen by the telescope but not the telescope+instrument. However, SGS measurements are not available for all AMUSING++ science observations, because not all FoVs contain a bright star suitable for SGS PSF measurement. We therefore use SGS FWHM measurements where available, and those from TEL IA FWHMLINOBS otherwise. When we must rely on the latter source, we correct for systematic offsets between it and the more accurate SGS measurement. \autoref{fig:fwhm} shows SGS FWHM versus TEL IA FWHMLINOBS measurements for each frame in AMUSING++. We perform a basic linear least squares fit between the two sets of measurements, the results of which are shown in the figure, and use this fit to convert the TEL IA FWHMLINOBS measurements to SGS FWHM.

There remains the question of how to assign an overall FWHM to the PSF of an observation that is a sum of multiple independent MUSE data frames, all of the same depth, but taken with different PSFs. For this purpose, we note that the dispersions of normal distributions average in quadrature, i.e., given a series of $I$ normal distributions $N(x,\sigma)$ centred on $x=0$ but with different dispersions $\sigma_i$, then the function $F = I^{-1} \sum_{i=1}^I N(0, \sigma_i)$ that is the average of these distributions has a variance given by $\sigma_F^2 = I^{-1}(\sigma_1^2 + \sigma_2^2 + \ldots \sigma_I^2)$. For this reason, we take the FWHM of an AMUSING++ image to be the root mean square of the FWHMs of the individual data frames from which it is assembled. Our pipeline also requires an uncertainty on this estimate, for the purposes of defining a prior on the PSF FWHM. We take this to be the standard deviation of the individual frame estimates. Thus, our final expression for the probability distribution of the FWHM for an AMUSING++ galaxy image is $p(\mathrm{FWHM}) = N(\mathrm{FWHM}_0,\mathrm{FWHM}_{\rm std})$, where
\begin{eqnarray}
    \mathrm{FWHM}_0 & = & \sqrt{\frac{1}{I} \sum_{i=1}^I \mathrm{FWHM}_i^2}
    \nonumber\\
    \mathrm{FWHM}_{\rm std} & = & \sqrt{\frac{1}{I}\sum_{i=1}^I (\mathrm{FWHM}_i - \mathrm{FWHM}_{0})^2}
\end{eqnarray}
are the root mean square and standard deviation of the PSF measurements $\mathrm{FWHM}_i$ for the $I$ individual frames from which it is assembled.

\section{Ionisation parameters in the D16 metallicity diagnostic}
\label{app:ion_par}

As discussed in the main text and suggested by \citet{Kewley19}, in principle our metallicities derived from the D16 diagnostic could be improved by making an independent estimate of the ionisation parameter $\log(U)$ using strong emission lines in MUSE. For optical data, the two diagnostics O32 and S32 are commonly used for this purpose. However, O32 uses the line ratio [O \textsc{iii}]$\lambda5007$ / [O \textsc{ii}]$\lambda\lambda3727,9$, and the lines appearing in the denominator are outside MUSE's wavelength coverage for nearby galaxies. S32 stands for ([S \textsc{iii}]$\lambda9069$ + [S \textsc{iii}]$\lambda9531$) / [S \textsc{ii}]$\lambda\lambda6717,31$, and [S \textsc{iii}]$\lambda9531$ is also outside MUSE's wavelength range. However, it could be estimated using a constant ratio 2.47 between [S \textsc{iii}]$\lambda9531$ and [S \textsc{iii}]$\lambda9069$ (which is within the MUSE window) as suggested both theoretically \citep{Vilchez96, Hudson12} and implemented in \textsc{pyneb} \citep{Luridiana15}. By adopting this ratio, one could estimate both metallicity and ionisation parameter at the same time using an iterative process \citep[e.g.][]{KD02}. However, after experimenting with this approach using both S32 maps or a single median value for each galactic field, we find that the typical S32 value is below 0. This finding is consistent with Figure 3 of \cite{Kreckel19}. Such a value corresponds to an unphysically low $\log(U) < -3.5$ and suggests that either the S32 diagnostic is failing for some reason, or that the assumed ratio of 2.47 for [S \textsc{iii}]$\lambda9531$/[S \textsc{iii}]$\lambda9069$ is not correct. Given the lines available in the MUSE wavelength range, it is not possible to decide between these possibilities, and other strong line calibrations of $\log(U)$ that can be computed from MUSE-accessible lines disagree with one another at the order of magnitude level \citep{Dors11}. For this reason, in this work we do not report any ionisation parameters, and adopt the original D16 metallicity diagnostic without ionisation parameter corrections.

\section{Pipeline validation tests}
\label{app:test}

\begin{figure}
\includegraphics[width=1.0\linewidth]{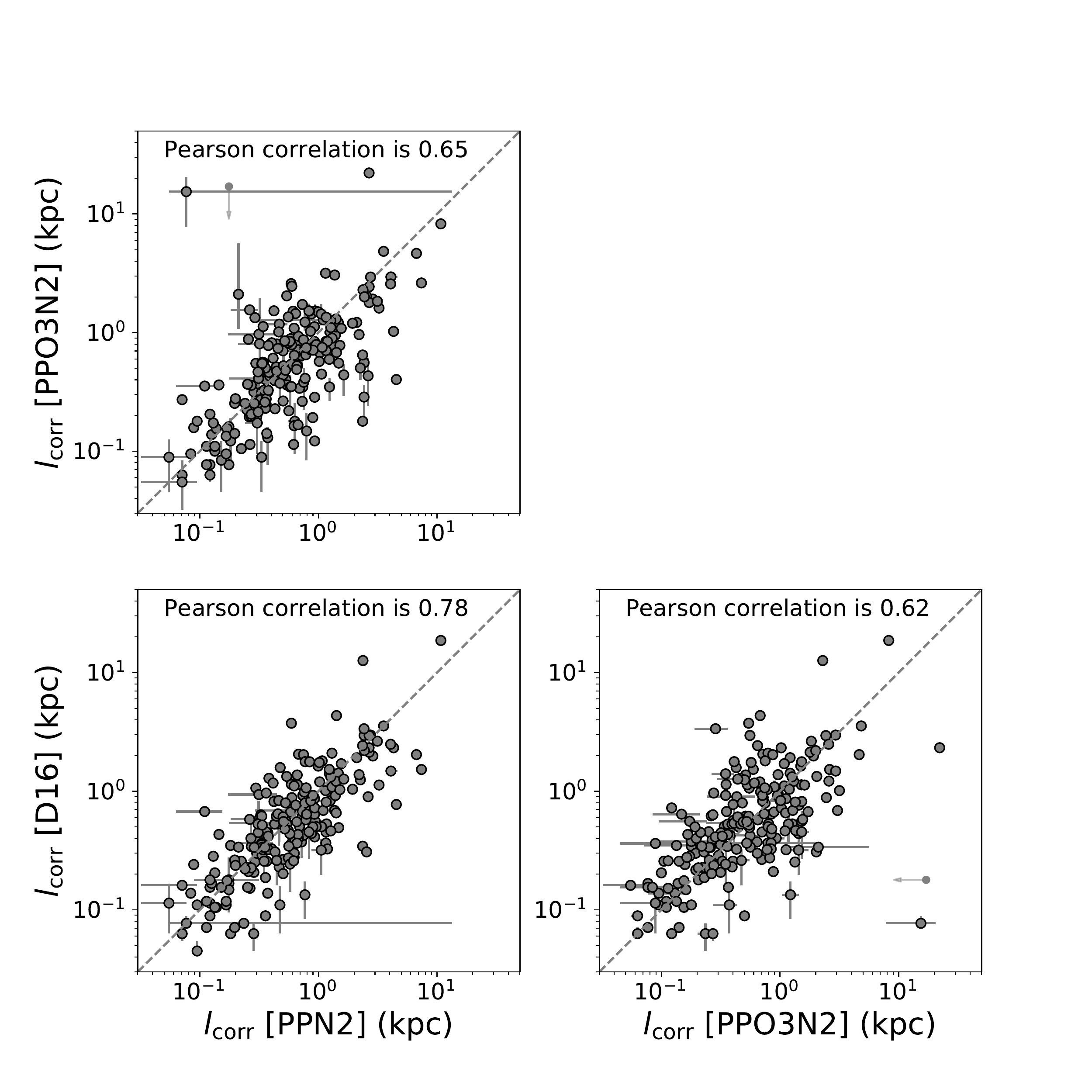}
\caption{Comparison of correlation length derived using three possible pairs of diagnostics: D16 (our fiducial choice), PPN2, and PPO3N2. Circles show the 50th percentile values for each galaxy, and error bars show the 16th to 84th percentile range.}
\label{fig:diag}
\end{figure}

\begin{figure}
\includegraphics[width=1.0\linewidth]{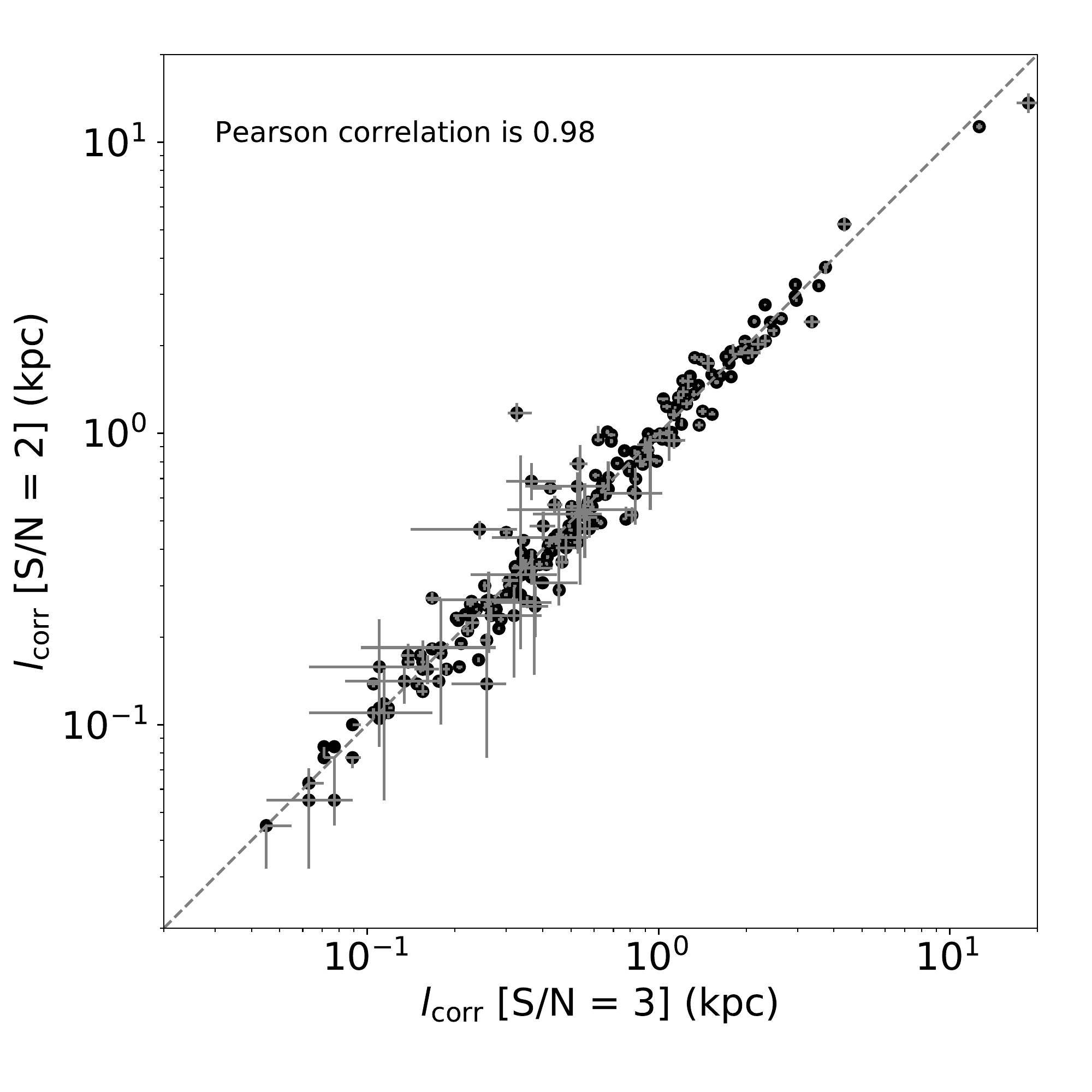}
\caption{Comparison of correlation length derived using S/N $=3$ versus S/N $=2$. Circles show the 50th percentile values for each galaxy, and error bars show the 16th to 84th percentile range. The diagonal grey dashed line shows the 1–1 line.}
\label{fig:sn}
\end{figure}

\begin{figure}
\includegraphics[width=1.0\linewidth]{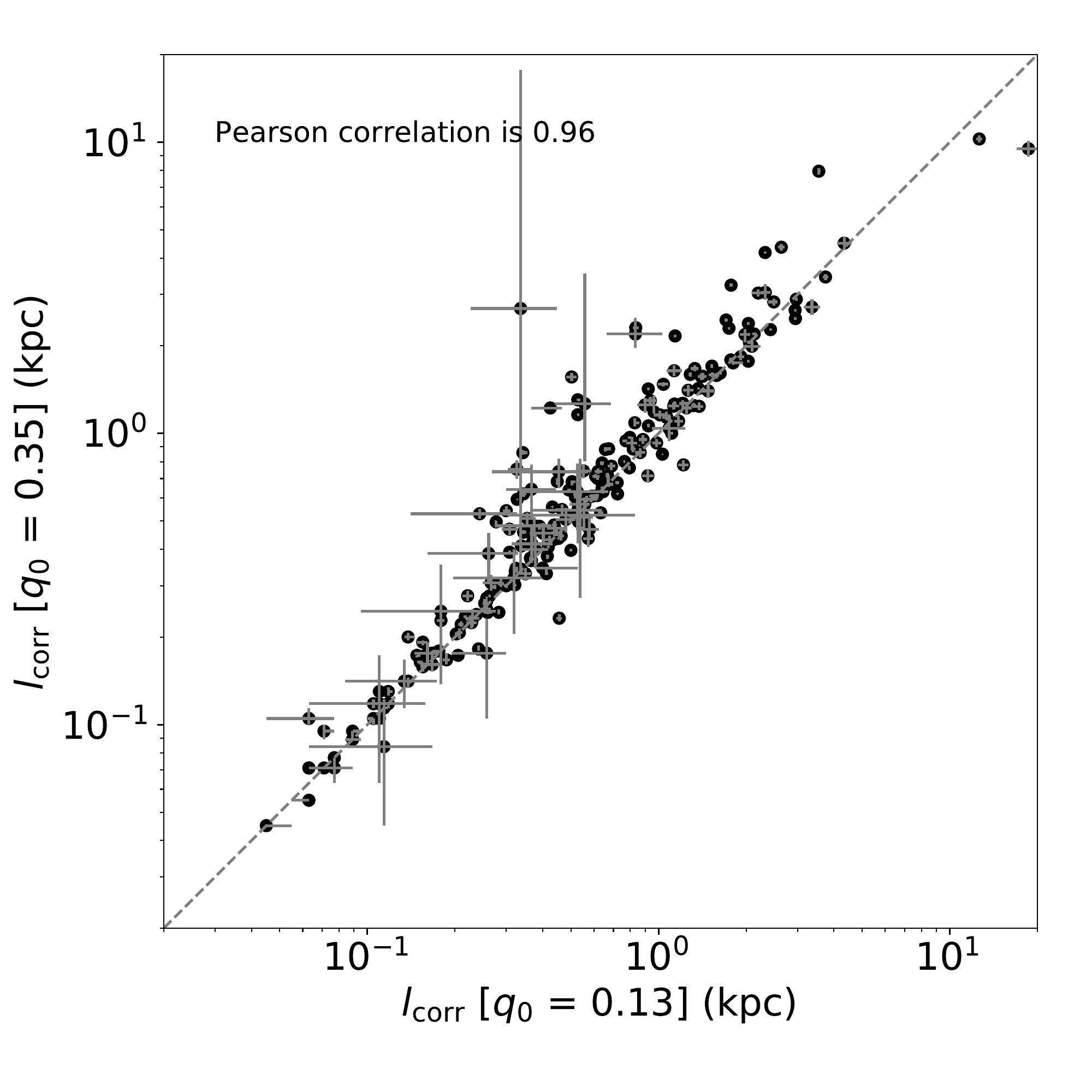}
\caption{Comparison of correlation length derived using $q_0 = 0.13$ (moderately thick discs) versus $q_0 = 0.35$ (very thick discs). Circles show the 50th percentile values for each galaxy, and error bars show the 16th to 84th percentile range.}
\label{fig:q0}
\end{figure}

\begin{figure}
\includegraphics[width=1.0\linewidth]{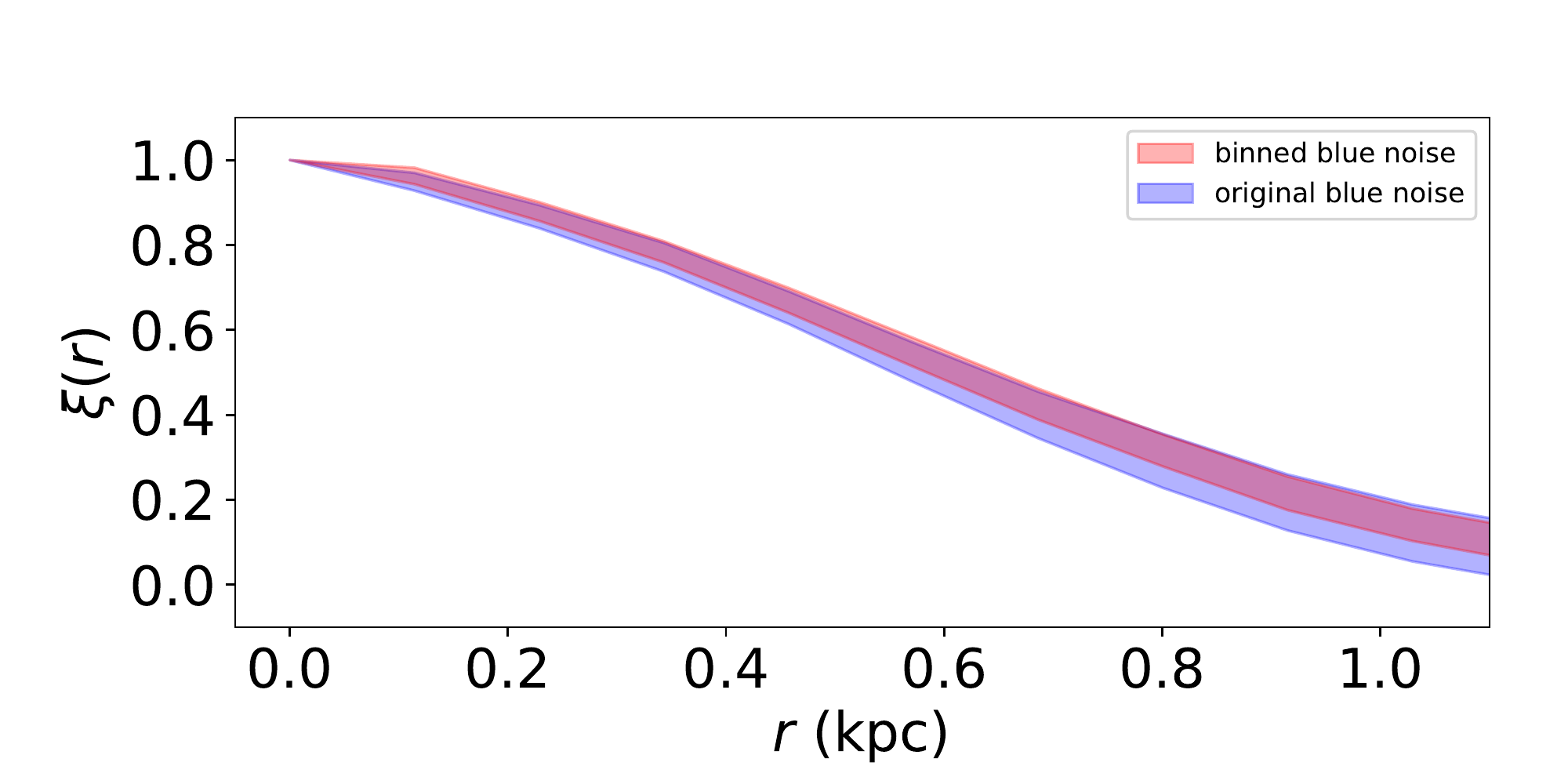}
\caption{The two-point correlation functions of an original noise map (the blue band) and a binned noise map (the red band) for NGC 7674.}
\label{fig:binned_noise}
\end{figure}

\begin{figure}
\includegraphics[width=1.0\linewidth]{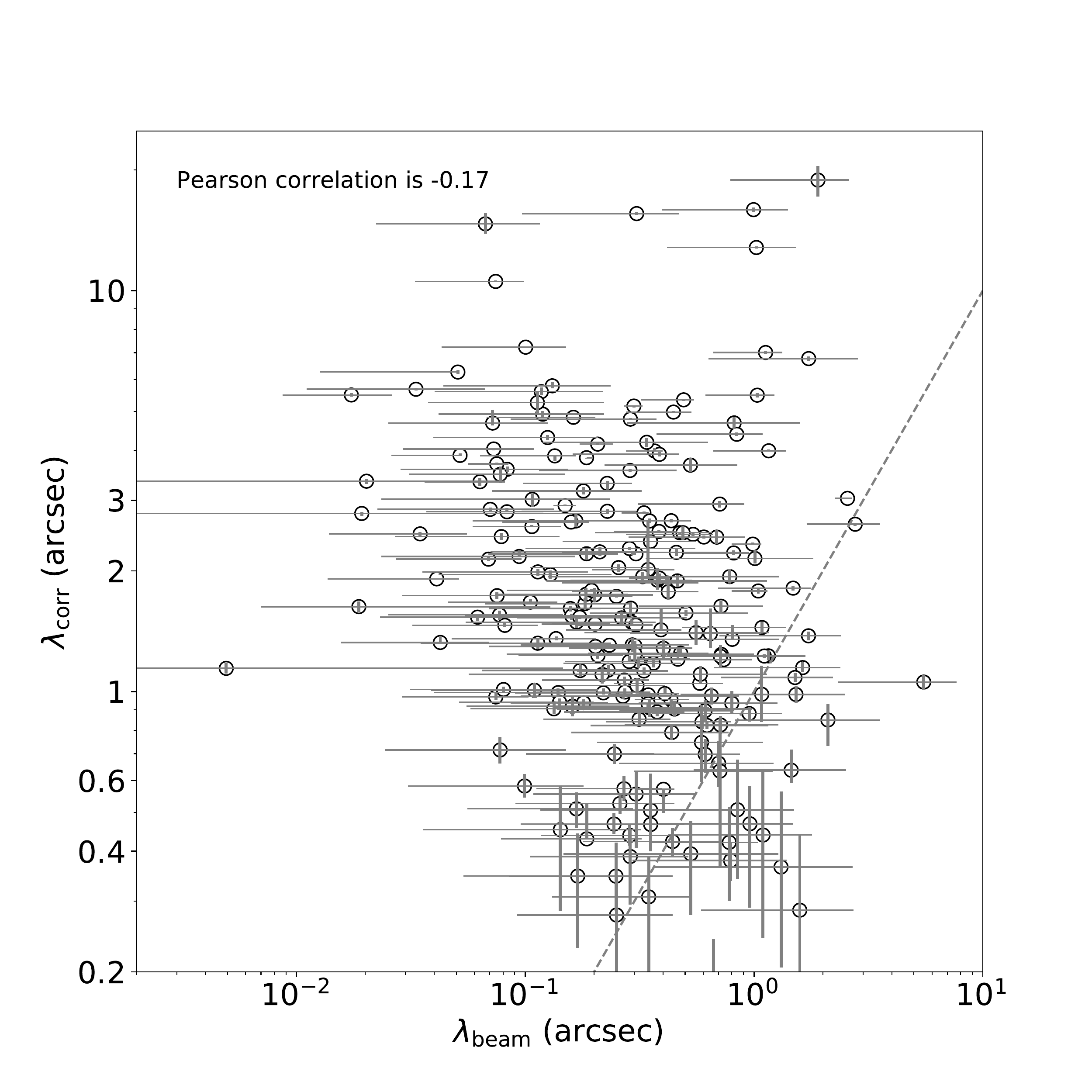}
\caption{The angular correlation length as a function of the angular semi major axis of the telescope beam projected on to the face of the galaxy. The grey dashed line is the 1–1 line.}
\label{fig:beam}
\end{figure}

\begin{figure*}
\includegraphics[width=1.0\linewidth]{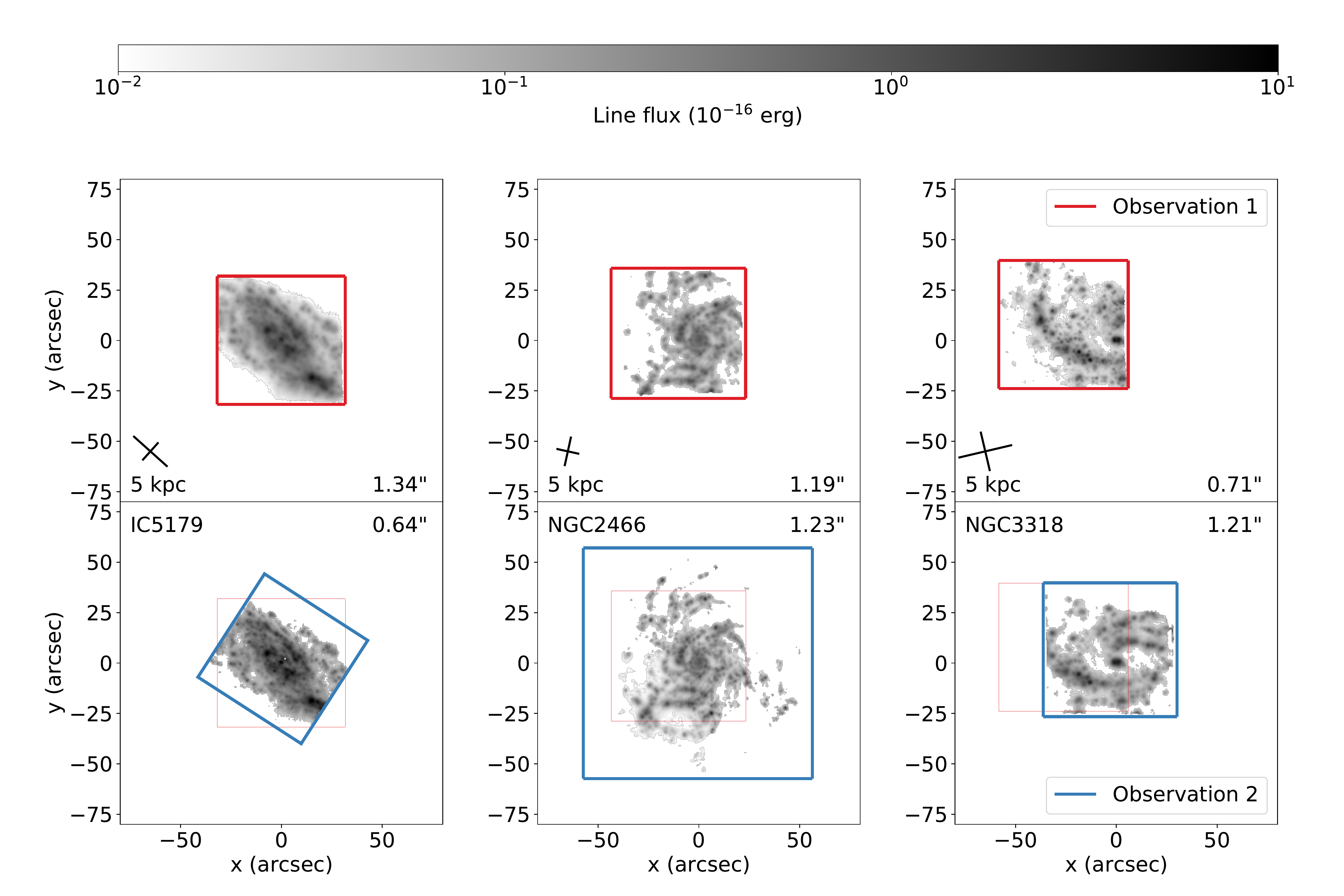}
\caption{H$\alpha$ line flux maps of IC5179 (Observation 1: IC+5179, upper left; Observation 2: target\_5\_centre, lower left), NGC 2466 (Observation 1: SN2016iye, upper middle; Observation 2: ASASSN14dd\_1, lower middle), and NGC 3318 (Observation 1: SN2000cl, upper right; Observation 2: SN2017ahn, lower right). In all panels, $x=y=0$ corresponds to the centre of the galaxy. The full coverage of Observation 1 is indicated using a red box and Observation 2 a blue box. The red boxes in the lower panels are identical to those in the upper panels, and for the purpose of illuminating the intersection regions. The cross scales are the same as \autoref{fig:large_lcorr}. The PSF of each observation is indicated in each panel.}
\label{fig:overlapping_image}
\end{figure*}

In this appendix we describe a series of validation tests we have performed on our pipeline. These tests are to verify that our results are not sensitive to the parameter choices that must be made, to search for spurious correlations that might indicate artificial resolution effects, and to check that our results are reproducible when details of the observations themselves are changed. We describe each of these tests below.

\subsection{Choice of metallicity diagnostic}
\label{app:diag}

Our first test, repeating one performed in \citetalias{L21}, is to compare our results derived using the D16 diagnostic to those derived from two other diagnostics, PPN2 and PPO3N2 \citep{PP04}, in order to ensure that are results are not dominated by systematic errors induced by the diagnostic. We derive metallicities from these two diagnostics and the rerun our full pipeline on the resulting metallicity maps. The results are shown in \autoref{fig:diag}. The corresponding Pearson correlation values for the diagnostic pairs PPN2-PPO3N2 (0.65), PPN2-D16 (0.78), and PPO3N2-D16 (0.62) confirm that the correlation lengths derived from different diagnostics show reasonable consistency.

\subsection{Signal-to-noise ratio}
\label{app:sn}

As discussed in \autoref{subsec:abr}, our adaptive binning reconstruction method requires that we choose a target S/N ratio and we adopt a minimum S/N of 3 for all analysis presented in the main text. This is a natural choice given that the S/N is not a significant bottleneck in recovering the low quality data using the adaptive binning algorithm. To verify that our results are not sensitive to this choice, we rerun our analysis pipeline using a target S/N ratio of 2 instead. \autoref{fig:sn} shows a comparison of $l_{\rm corr}$ derived using the two different S/N values. The results derived using the two S/N cuts cluster tightly around the 1-1 line. Quantitatively, the Pearson correlation between the 50th percentile values for the two sets of results is 0.98. This shows that changing the target S/N does not significantly influence the value of $l_{\rm corr}$ we derive.

\subsection{Intrinsic galaxy aspect ratio $q_0$}
\label{app:q0}

As discussed in \autoref{subsec:backbone}, deprojecting galaxy images requires an estimate of the intrinsic axis ratio $q_0$ in the original image. For all the analyses presented in the main text, we adopt $q_0=0.13$, a generally adopted value of moderate disc thickness. Given that our samples include various Hubble types (from S0 to I), however, it is important that we check if our derived values of $l_{\rm corr}$ depend on $q_0$, since it is possible that the intrinsic thicknesses of the galaxies in our sample vary systematically. \autoref{fig:q0} shows the comparison of $l_{\rm corr}$ derived using our fiducial choice $q_0=0.13$, and values derived from the same pipeline but assuming $q_0=0.35$, corresponding to a much thicker disc. The Figure shows that changing $q_0$ does not significantly influence $l_{\rm corr}$; the Pearson correlation between the data sets derived using the two different $q_0$ values is 0.96. Thus, we adopt a fixed $q_0 = 0.13$ in our analysis.

\subsection{Effect of adaptive binning on the two point correlation function}
\label{app:binning}

As discussed in \citetalias{L21} and \citetalias{KT18}, the finite resolution of an observation will induce an apparent correlation even in a completely uncorrelated map of noise. Measured two-point correlations are significant only to the extent that they are well above the floor induced by this effect. Our adaptive binning algorithm could in principle introduce additional spurious correlations on top of the telescope beam effect, and it is therefore important to characterise the extent to which it does so.

We perform this test on NGC 7674, which is fairly typical in terms of mass [$\log(M_*/$M$_{\odot})=11.39$] and star formation rate [$\log($SFR/M$_{\odot}$ yr$^{-1})=1.14$] in our sample, but is one of the more distant galaxies ($D=117.93$ Mpc) and thus the most strongly affected by beam smearing. For this galaxy, we construct a pure, uncorrelated noise map with the same resolution as the real map, and set the metallicity fluctuation in each pixel to a Gaussian random value with zero mean and unit variance. We then convolve the noise map with a Gaussian beam with the same FWHM that NGC 7674 has and apply the same pixel mask as for the original map. We then compute the two-point correlation function of the noise map using the same pipeline we apply to the real map, but without the adaptive binning step. We repeat this procedure 50 times and measure the mean and standard deviation of the resulting correlations. We plot the mean and a range of one standard deviation around it as the blue band in \autoref{fig:binned_noise}. This represents the noise imposed simply by the telescope beam.

To check the effects of the adaptive binning algorithm, we use the same procedure, except that after convolution of the noise map with the telescope beam, we adaptively bin the map. We do so using the same adaptive binning as we used for the actual NGC 7674 observations, and apply the same pixel mask as for the binned map. Thus, we have a reconstructed noise map. We then derive the two point correlation function as before, again repeating the procedure 50 times and computing the mean and standard deviation. We plot these as the red band in \autoref{fig:binned_noise}.

Comparison of the red and blue bands clearly indicates that adaptive binning only very slightly increases the correlation floor. Quantitatively, applying our parametric fitting pipeline to the data shown by the red and blue bands results in values of $l_{\rm corr}$ and $w_{\rm inj}$ that are both close to zero, and with means that differ between the two cases by $\lesssim 25$ pc. For comparison, our best-fit values of $l_{\rm corr}$ are almost always much larger than 25 pc, which suggests that adaptive binning has no significant effects on our estimates of it. By contrast, our estimates of $w_{\rm inj}$ are not necessarily very large compared to this value, which is why we derive $w_{\rm inj}$ from maps that is original and have not been adaptively binned.

\subsection{Artificial correlations due to distance-dependent spatial resolution}

\begin{figure}
\includegraphics[width=1.0\linewidth]{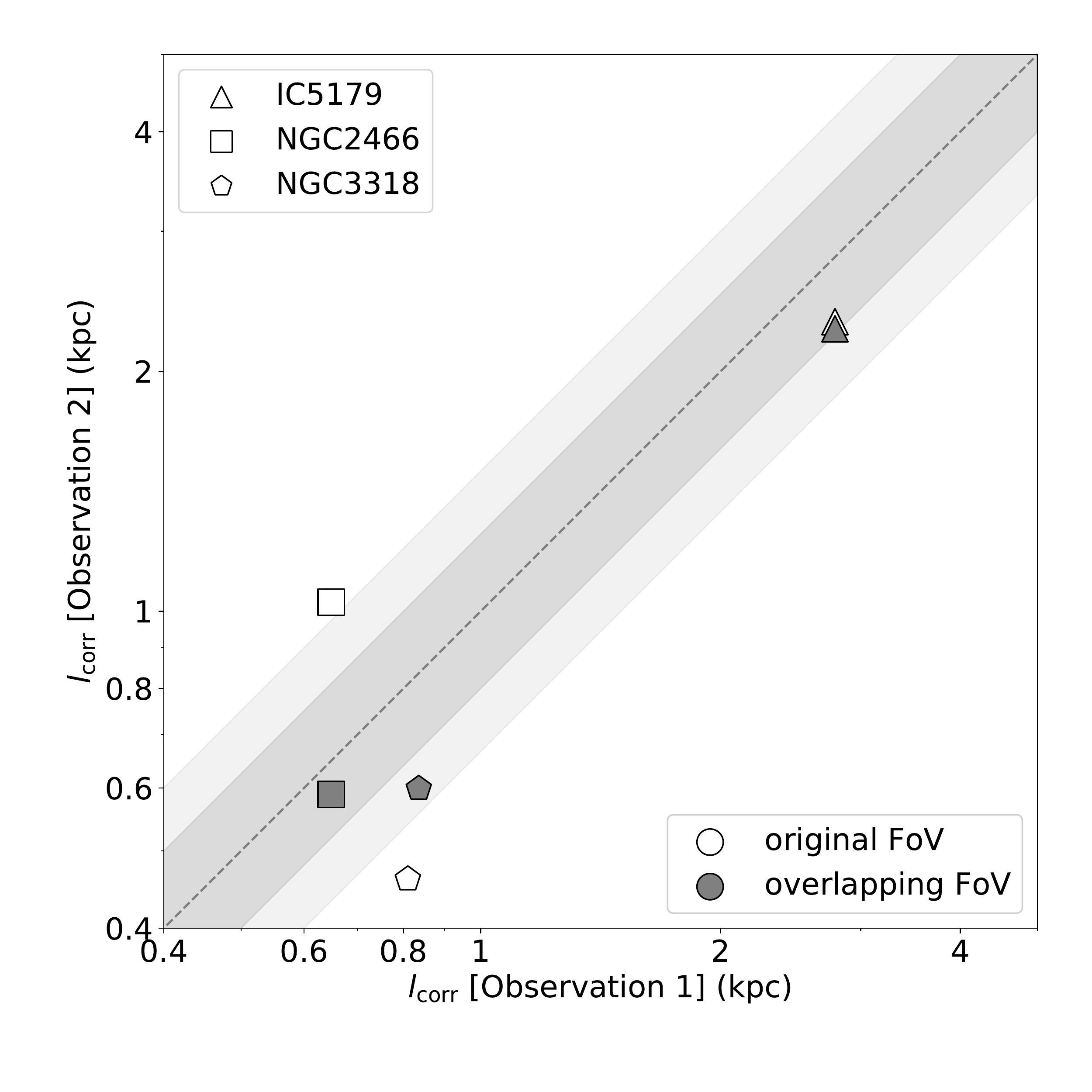}
\caption{Comparison of correlation lengths of the same galaxy derived using two separate observations. Triangles, squares, and pentagons show IC5179, NGC 2466, and NGC 3318, respectively. Open markers show the 50th percentile values of $l_{\rm corr}$ for the original FoV, and closed markers show the overlapped FoV. The grey and light grey bands show the 25\% and 50\% difference range from the dashed 1-1 line. In some cases, the common area between two observations is Observation 1 itself (see \autoref{fig:overlapping_image}), so $l_{\rm corr}$ does not move horizontally along the $x$-axis. Error bars showing the 16th to 84th percentile range are not included, since they are smaller than the plot symbols and will not be visible.}
\label{fig:overlapping_lcorr}
\end{figure}

Our next test is reproduced from \citetalias{L21}, and checks for spurious correlations induced by the fact that the physical size of the telescope beam, projected on the galaxy, is a function of galactic distance. This provides a check of whether the correlation lengths we are measuring are robust as opposed to being an artifact of the finite resolution of the observations: if they are real, there should be no systematic relationship between beam size and measured correlation length, while if they are spurious there should be such a correlation. For the purpose of this test, we define the projection-corrected angular beam size to be
\begin{equation}
\lambda_{\rm beam} = \frac{\sigma_{\rm beam,a}}{\cos i},
\end{equation}
and the angular correlation length to be
\begin{equation}
\lambda_{\rm corr} = \frac{l_{\rm corr}}{D},
\end{equation}
where $\sigma_{\rm beam,a}$ is the angular beam size from the posteriors and $\cos i$ is computed from \autoref{eqn:cosi}. We show $\lambda_{\rm corr}$ as a function of $\lambda_{\rm beam}$ in \autoref{fig:beam}. Visual inspection shows no evidence of significant correlation, and the Pearson correlation of -0.17 that we find from the data is consistent with this visual impression. This test suggests that the correlations lengths we measure are real measurements, and are not simply artifacts of the spatial resolution of the maps.

\subsection{Repeated observations}

Our final test of the robustness of our pipeline is to take advantage of the AMUSING++ compilation, which collects all the galaxies observed by MUSE. As a result, it contains three galaxies that were observed twice and tagged with different names. They are IC5179 (Observation 1: IC+5179, Observation 2: target\_5\_centre), NGC 2466 (Observation 1: SN2016iye, Observation 2: ASASSN14dd\_1), and NGC 3318 (Observation 1: SN2000cl, Observation 2: SN2017ahn). Their H$\alpha$ maps are shown in \autoref{fig:overlapping_image}.

Because the motivations for the two observations were different, the two observations differ not just in the orientation of the galaxy in the field of view, but also in which part of the galaxy is captured within the FoV (in some cases the full coverage is segmented by multiple FoVs), in the depth of the observation, and in the seeing conditions. This provides a unique opportunity to test the robustness of our analysis. We do so by processing the paired observations independently through our pipeline. We do this in two ways. First, we simply analyse each observation separately, as if it were not part of a pair. The advantage of this is that, since the fields of view are not exactly the same, this tests the robustness of our analysis against differences in what part of the target galaxy is captured within the field of view. The disadvantage of this approach is that it is possible that there is a \textit{real} difference in correlation lengths between the  parts of the galaxy captured by the two fields, and thus even if our pipeline produces different answers for the two fields, it might be detecting a real difference. For this reason, we also perform a second comparison in which we crop the fields of view of both observations to include only the region where they overlap, and we are therefore analysing images of the same patch of the galaxy, but observed with different depth and seeing.

We show a scatter plot of $l_{\rm corr}$ versus $l_{\rm corr}$ in \autoref{fig:overlapping_lcorr}, taking Observation 1 as $x$-axis and Observation 2 $y$-axis; open symbols show the analysis using the original field of view, while filled symbols show the results using only the overlapping parts of the field of view. We find good agreement, with all three cases using overlapping fields of view falling within $\approx 25\%$ of the 1-1 line. As expected, comparison between open (original FoV) and closed (overlapped FoV) markers demonstrates that $l_{\rm corr}$ from the overlapping fields is closer to the 1-1 line than in the cases when the fields differ, but the difference is relatively small. In either case, this test provides confidence regarding our whole pipeline.

%%%%%%%%%%%%%%%%%%%%%%%%%%%%%%%%%%%%%%%%%%%%%%%%%%

% Don't change these lines
\bsp	% typesetting comment
\label{lastpage}
\end{document}